%% file: ggFVBF2017Paper.tex
\pdfoutput=1
\newcommand*{\ATLASLATEXPATH}{latex/}
\documentclass[cernpreprint, texlive=2016, UKenglish]{\ATLASLATEXPATH atlasdoc}

\usepackage{\ATLASLATEXPATH atlaspackage}
\usepackage{\ATLASLATEXPATH atlasbiblatex}

\usepackage{\ATLASLATEXPATH atlasphysics}
\usepackage{multirow}
\usepackage{float}
\usepackage{feynmf}
\usepackage{pgfplots}
\usepackage{subfig}

\graphicspath{{logos/}{figures/}}

\usepackage{ggFVBF2017Paper-defs}

\FloatBarrier

\input{ggFVBF2017Paper-metadata}

\hypersetup{pdftitle={ATLAS document},pdfauthor={The ATLAS Collaboration}}

\usepackage{graphicx}

\begin{document}

\maketitle

\input{sections/PaperOutline}
\printbibliography

\clearpage
 \input{atlas_authlist}

\clearpage



\end{document}

%% file: ggFVBF2017Paper-metadata.tex
\AtlasTitle{Measurements of gluon--gluon fusion and vector-boson fusion Higgs boson production cross-sections in the $\hwwenmn$ decay channel in $pp$ collisions at $\sqrt{s}=13~\TeV$ with the ATLAS detector}

\AtlasAbstract{%
Higgs boson production cross-sections in proton--proton collisions are measured in the $\hwwenmn$ decay channel. 
The proton--proton collision data were produced at the Large Hadron Collider at a centre-of-mass energy of 13~\TeV\ and recorded by the ATLAS detector in 2015 and 2016, corresponding to an integrated luminosity of 36.1~fb$^{-1}$.
The product of the $\hww$ branching fraction times the gluon--gluon fusion and vector-boson fusion cross-sections are measured to be 
$11.4 ^{+1.2}_{-1.1} (\mathrm{stat.}) ^{+1.8}_{-1.7}(\mathrm{syst.})~\mathrm{pb}$ and $0.50 ^{+0.24}_{-0.22} (\mathrm{stat.}) \pm 0.17(\mathrm{syst.})~\mathrm{pb}$, 
respectively, in agreement with Standard Model predictions.
}

\AtlasRefCode{HIGG-2016-07}
\PreprintIdNumber{CERN-EP-2018-212}

\AtlasJournal{Phys.\ Lett.\ B.}
\AtlasJournalRef{\PLB 789 (2019) 508}
\AtlasDOI{10.1016/j.physletb.2018.11.064}

\AtlasCoverCommentsDeadline{31 July 2018}

\AtlasCoverEgroupEditors{atlas-higg-2016-07-editors@cern.ch}

\AtlasCoverEgroupEdBoard{atlas-higg-2016-07-editorial-board@cern.ch}

%% file: sections/PaperOutline.tex
\section{Introduction}
\label{sec:intro}
\input{sections/Intro_Letter}

\section{ATLAS detector}
\label{sec:detector}
\input{sections/ATLAS_Letter}

\section{Signal and background Monte Carlo predictions}
\label{sec:MCData}
\input{sections/MC_Letter}

\section{Event selection and categorisations}
\label{sec:EventSelection}
\input{sections/EvtSelection_Letter}

\section{Background estimation}
\label{sec:Background}
\input{sections/Background_Letter}


\section{Systematic uncertainties}
\label{sec:SystematicUncertainties}
\input{sections/Systematics_Letter}

\section{Signal region yields and results}
\label{sec:Result}
\input{sections/SignalExtraction_Letter}
\input{sections/Results_Letter}


\section{Conclusions}
\label{sec:Conclusions}
\input{sections/Conclusions_Letter}

\section{Acknowledgements}
\label{sec:Acknowledgements}
\input{sections/Acknowledgements}

%% file: sections/Intro_Letter.tex
This Letter presents a measurement of the inclusive Higgs boson production cross-sections via gluon--gluon fusion (ggF) and 
vector-boson fusion (VBF) through the decay $\hwwenmn$ using 36.1~fb$^{-1}$ of proton--proton  collisions 
at a centre-of-mass energy of 13~\TeV\ recorded by the ATLAS detector. 
Higgs boson couplings have been studied in this channel with Run-1 data by the ATLAS~\cite{HIGG-2013-13} and CMS~\cite{CMS-HIG-12-036} experiments and recently with Run-2 data by the CMS experiment~\cite{Sirunyan:2018egh}.
The $\hww$ decay channel has the second-largest branching fraction 
and allowed the most precise Higgs boson cross-section measurements in Run-1~\cite{HIGG-2015-07}. 
The measured cross-section of the ggF production process probes the Higgs boson couplings to 
gluons and heavy quarks, while the VBF process directly probes the couplings to $W$ and $Z$ bosons.
The leading-order diagrams for the ggF and VBF production processes are depicted in Figure~\ref{fig:production}.

\input{figures/fig_production}

%% file: figures/fig_production.tex
\begin{figure}[!h]
\centering
\raisebox{-0.5\height}{
\captionsetup[subfloat]{captionskip=30pt}
\subfloat[ggF production]{
\begin{fmffile}{ggFgraph}
  \begin{fmfgraph*}(170,60)
    \fmfstraight
    \fmfleft{i1,i2}
    \fmfright{o1,m,o2}
    \fmf{gluon}{i1,t1}
    \fmf{gluon}{t2,i2}
    \fmflabel{$g$}{i1}
    \fmflabel{$g$}{i2}
    \fmf{phantom,tension=0.4}{t1,o1}
    \fmf{phantom,tension=0.4}{t2,o2}
    \fmffreeze
    \fmf{plain,tension=1}{t1,t2,t3,t1}
    \fmf{phantom,tension=1.4}{t3,m}
    \fmffreeze
    \fmf{dashes,tension=1.4,label=H,label.side=left}{t3,h}
    \fmf{photon,tension=1}{h,o1}
    \fmf{photon,tension=1}{h,o2}
    \fmfblob{80}{h}
    \fmfv{decor.shape=circle,decor.filled=empty,decor.size=80}{t3}
    \fmflabel{$W^{*}$}{o1}
    \fmflabel{$W$}{o2}
  \end{fmfgraph*}
\end{fmffile}
}
}
\hspace{1cm}
\raisebox{-0.5\height}{
\captionsetup[subfloat]{captionskip=15pt}
\subfloat[VBF production]{
\begin{fmffile}{VBFgraph}
 \begin{fmfgraph*}(150,90)
    \fmfleft{d,i2,d,d,i1,d}
    \fmfright{q2,t,b,q1}
    \fmftop{top}
    \fmfbottom{bot}
    \fmf{plain,tension=1.9}{i1,v1}
    \fmf{plain,tension=1.9}{i2,v2}
    \fmf{plain}{v1,q1}
    \fmf{plain}{q2,v2}
    \fmf{boson,label=$V$,label.side=right}{v1,h,v2}
    \fmf{dashes,tension=1.1,label=H,label.side=left}{h,a}
    \fmf{photon,label.side=right}{b,a,t}
    \fmfblob{80}{a}
    \fmfblob{80}{h}
    \fmflabel{$W^{*}$}{t}
    \fmflabel{$W$}{b}
    \fmflabel{$q$}{i1}
    \fmflabel{$q$}{i2}
    \fmflabel{$q'$}{q1}
    \fmflabel{$q'$}{q2}
    \fmf{phantom,tension=0.4}{top,b,t,bot}
  \end{fmfgraph*}
\end{fmffile}
\vspace{10cm}
}
}

  \caption{
      Diagrams for the leading production modes (ggF and VBF), where the
      $VV\!H$ and $qqH$ coupling vertices are marked with shaded and empty circles, respectively. 
      The $V$ represents a $W$ or $Z$ vector boson.
    }
  \label{fig:production}
\end{figure}
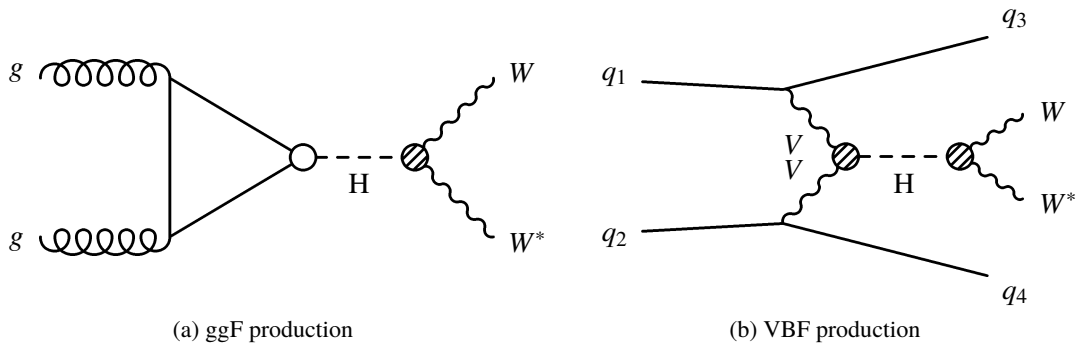

%% file: sections/ATLAS_Letter.tex
ATLAS is a particle detector designed to achieve a nearly full coverage in solid angle\footnote{ATLAS uses a right-handed coordinate system with its origin at the nominal interaction point (IP) in the centre of the detector and the $z$-axis along the beam pipe. The $x$-axis points from the IP to the centre of the LHC ring, and the $y$-axis points upward. Cylindrical coordinates $(r, \phi)$ are used in the transverse plane, $\phi$ being the azimuthal angle around the $z$-axis. The pseudorapidity is defined in terms of the polar angle $\theta$ as $\eta = -\ln\tan(\theta/2)$. The distance in ($\eta$,$\phi$) coordinates, $\Delta R = \sqrt {\Delta \phi^2+ \Delta \eta^2}$, is also used to define cone sizes. Transverse momentum and energy are defined as $\pT = p \sin \theta$ and $\et = E \sin \theta$, respectively.}~\cite{PERF-2007-01,CERN-LHCC-2012-009}.
It consists of an inner tracking detector surrounded by a thin superconducting solenoid, electromagnetic and hadronic calorimeters,
and a muon spectrometer incorporating three large superconducting air-core toroidal magnets.
The inner tracking detector (ID) is located in a 2 T magnetic field and is designed to measure charged-particle
trajectories up to a pseudorapidity of $|\eta| = 2.5$.
Surrounding the ID are electromagnetic and hadronic calorimeters, which
use liquid argon (LAr) and lead absorber for the electromagnetic central and endcap calorimeters ($|\eta| < 3.2$), copper absorber for the hadronic endcap calorimeter ($1.5 < |\eta| < 3.2$), 
and scintillator-tile active material with steel absorber for the central ($|\eta| < 1.7$)
hadronic calorimeter. The solid angle coverage is extended to $|\eta| = 4.9$ with forward copper/LAr and tungsten/LAr calorimeter modules.
The muon spectrometer comprises separate trigger chambers within the range $|\eta| < 2.4$ and
high-precision tracking chambers within the range $|\eta| < 2.7$, measuring the deflection of muons in a magnetic field generated by the three superconducting toroidal magnets.
A two-level trigger system is used to select events~\cite{Aaboud:2016leb}.

%% file: sections/MC_Letter.tex
Higgs boson production via ggF was simulated at next-to-next-to-leading-order (NNLO) accuracy in QCD using the \POWHEGBOXV{v2} NNLOPS program~\cite{Hamilton:2013fea}, 
with the PDF4LHC15 NNLO set of parton distribution functions (PDF)~\cite{0954-3899-43-2-023001}. 
The simulation achieves NNLO accuracy for arbitrary inclusive $gg \to H$ observables by reweighting the Higgs boson rapidity spectrum 
in Hj-MiNLO~\cite{Campbell:2012am} to that of HNNLO~\cite{Catani:2007vq}. 
The transverse momentum spectrum of the Higgs boson obtained with this sample was found to be compatible within uncertainties with the resummed NNLO+NNLL HRes2.3 calculation~\cite{Bozzi:2005wk,deFlorian:2011xf}.
The parton-level events produced by the \POWHEGBOXV{v2} NNLOPS program were passed to \PYTHIAV{8}~\cite{SJOSTRAND2008852} to provide
parton showering, hadronisation and the underlying event, using the AZNLO set of data-tuned parameters~\cite{STDM-2012-23}.

Higgs boson production via VBF was simulated at next-to-leading-order~(NLO) accuracy in QCD using~\POWHEGBOXV{v2}~\cite{Nason2010,Hamilton:2013fea,Hamilton:2015nsa,Campbell:2012am} with the PDF4LHC15 NLO PDF set~\cite{0954-3899-43-2-023001}.
The parton-level events were passed to \PYTHIAV{8}~\cite{SJOSTRAND2008852} with the same parameters as for ggF.

The mass of the Higgs boson was set to 125~\GeV, compatible  with the experimental measurement~\cite{HIGG-2014-14,HIGG-2016-33,CMS-HIG-16-041}. 
The corresponding Standard Model (SM) branching fraction $\mathcal{B}_{\hww}$ is calculated using HDecay v6.50~\cite{DJOUADI199856,Djouadi:2018xqq} to be 0.214~\cite{deFlorian:2016spz}. 
The \hwwlnln decay, where $\ell=e$ or $\mu$, always includes the small contribution from $W \to \tau \nu  \to \ell \nu \nu \nu$ decays.
Other production and decay modes of the Higgs boson are either fixed to SM predictions ($VH$ production and $\htautau$ decay) or neglected ($t\bar{t}H$ and $b\bar{b}H$ associated production).

The ggF production cross-section was calculated with next-to-next-to-next-to-leading-order 
accuracy in QCD and includes NLO electroweak (EW) corrections~\cite{Heinemeyer:1559921,PhysRevLett.114.212001,Anastasiou2016,ACTIS200812,1126-6708-2009-04-003}. The NLO QCD and EW calculations are used with approximate NNLO QCD corrections for the VBF production cross-section~\cite{Heinemeyer:1559921,PhysRevLett.99.161803,PhysRevD.77.013002,PhysRevLett.105.011801}.

The $WW$ background was generated separately for the $qq \to WW$ and $gg \to WW$ production mechanisms. 
The $qq \to WW$ production process was generated using \SHERPAV{2.2.2}~\cite{Gleisberg:2008ta,Gleisberg:2008ta,Gleisberg:2008fv} interfaced with the NNPDF3.0 NNLO PDF set~\cite{Ball:2014uwa} and the \SHERPA parton shower, hadronisation and underlying event simulation (UEPS) model~\cite{1126-6708-2008-03-038,Hche2013}. 
The matrix elements were calculated for up to one additional parton at NLO and up to three additional partons at LO precision.
The loop-induced $gg \to WW$  process was simulated by \SHERPAV{2.1.1} with zero or one additional jet~\cite{Cascioli:2013gfa}.
The sample is normalised to the NLO $gg \to WW$ cross-section~\cite{Caola:2015rqy}.
Interferences with direct $W$W production have a negligible impact after event selection cuts have been applied and are, therefore, not considered in this analysis \cite{Campbell:2011cu}.

While NNLO cross-sections are available for diboson production processes~\cite{Grazzini:2016ctr,Grazzini:2017ckn,Cascioli:2014yka}, 
the \SHERPA \break MEPS@NLO prescription~\cite{Hche2013} is used in this analysis. 
This procedure already captures the majority of the NNLO shape corrections.

The MC generators, PDFs, and programmes used for the UEPS are summarised in Table~\ref{tab:mcsamples}. 
The order of the perturbative prediction for each sample is also reported.

The generated events were passed through a \GEANT~4~\cite{AGOSTINELLI2003250} simulation of the ATLAS detector~\cite{Aad:2010ah} 
and reconstructed with the same analysis software as used for the data.
Additional proton–proton interactions (pile-up) are included in the simulation for all generated events
such that the distributions of the average number of interactions per bunch crossing reproduces that observed in the data.
The inelastic proton–proton collisions were produced using \PYTHIAV{8} with the A2 set of data-tuned 
parameters~\cite{ATL-PHYS-PUB-2012-003} and the MSTW2008LO PDF set~\cite{Martin:2009iq}. 
Correction factors are applied to account for small differences observed between data and simulation in electrons, muons, and jets identification efficiencies and energy/momentum scales and resolutions.

\begin{table}[h]
  \centering
  \caption{Overview of simulation tools used to generate signal and background processes, and to model the UEPS. The PDF sets are also summarised. 
Alternative event generators and configurations used to estimate systematic uncertainties are shown in parentheses.}
  \label{tab:mcsamples}
\scalebox{0.75}{
  \begin{tabular}{l l l l l}
  \hline\hline
  Process              & Matrix element                      & PDF set & UEPS model                           & Prediction order\\
                       & (alternative)                       &      & (alternative model)                &  for total cross-section\\
  \hline
  ggF $H$                 & \POWHEGBOXV{v2}                                                  & \multirow{2}{*}{PDF4LHC15 NNLO~\cite{0954-3899-43-2-023001}}       & \multirow{2}{*}{\PYTHIAV{8}~\cite{SJOSTRAND2008852}} & \multirow{2}{*}{N$^{3}$LO QCD + NLO EW~\cite{Heinemeyer:1559921,PhysRevLett.114.212001,Anastasiou2016,ACTIS200812,1126-6708-2009-04-003}}\\
                       & NNLOPS~\cite{Nason2010,Hamilton:2013fea,Campbell:2012am}     &                                                   &                                      &  \\
                       & (\MGFiveNLO~\cite{Alwall:2014hca,Frederix:2012ps})&                  & (\HerwigV{7}~\cite{Bellm:2015jjp}) & \\     
  VBF $H$                 & \POWHEGBOXV{v2}            & PDF4LHC15 NLO                     & \PYTHIAV{8} & NNLO QCD + NLO EW~\cite{Heinemeyer:1559921,PhysRevLett.99.161803,PhysRevD.77.013002,PhysRevLett.105.011801}\\
                       &            &                             & (\HerwigV{7}) & \\   
   $VH$                &  \POWHEGBOXV{v2}~\cite{Luisoni2013}            & PDF4LHC15 NLO                     & \PYTHIAV{8} & NNLO QCD + NLO EW\cite{HanWillenbrock1991,BreinDjouadiHarlander2004,CiccoliniDittmaier2003}\\
  $qq \to WW$          & \SHERPAV{2.2.2}~\cite{Gleisberg:2008ta,Gleisberg:2008ta,Gleisberg:2008fv}    & NNPDF3.0NNLO~\cite{Ball:2014uwa} & \SHERPAV{2.2.2}~\cite{1126-6708-2008-03-038,Hche2013} & NLO~\cite{Cascioli:2013gfa} \\ 
                       & (\POWHEGBOXV{v2},               &                          & \multirow{2}{*}{(\Herwigpp~\cite{Bellm:2015jjp})}  \\
                       & \MGFiveNLO)                 &                          &  \\
$gg \to WW$            & \SHERPAV{2.1.1}~\cite{Cascioli:2013gfa}& CT10~\cite{Lai:2010vv}   & \SHERPAV{2.1}                 & NLO~\cite{Caola:2015rqy}  \\
$WZ/V\gamma^{\ast}/ZZ$ & \SHERPAV{2.1}                         & CT10                     & \SHERPAV{2.1}                  & NLO~\cite{Cascioli:2013gfa} \\
  $V\gamma$            & \SHERPAV{2.2.2}                    & NNPDF3.0NNLO             &\SHERPAV{2.2.2}            & NLO~\cite{Cascioli:2013gfa} \\
                       & (\MGFiveNLO)                        &                          & (CSS variation~\cite{1126-6708-2008-03-038,Hoeche:2009xc}) &\\
  $t\bar{t}$           & \POWHEGBOXV{v2}~\cite{Frixione:2007nw}  & NNPDF3.0NLO       &\PYTHIAV{8} & NNLO+NNLL~\cite{Czakon:2013} \\  
                       & (\SHERPAV{2.2.1} )                     &                            & (\HerwigV{7})   & \\
 $Wt$                  & \POWHEGBOXV{v1}~\cite{Re:2010bp}       & CT10~\cite{Lai:2010vv}           &  \PYTHIAV{6.428}~\cite{Sjostrand:2006za} & NLO~\cite{Re:2010bp}                \\
                       & (\MGFiveNLO)                       &                                  & (\Herwigpp)    &     \\
  $Z/\gamma^{\ast}$    & \SHERPAV{2.2.1}                      & NNPDF3.0NNLO                     & \SHERPAV{2.2.1}         & NNLO~\cite{Melnikov:2006kv,Anastasiou:2003ds} \\
\hline\hline
  \end{tabular}
}
\end{table}

%% file: sections/EvtSelection_Letter.tex
\begin{figure}[!h]
  \centering
    \includegraphics[width=0.55\textwidth]{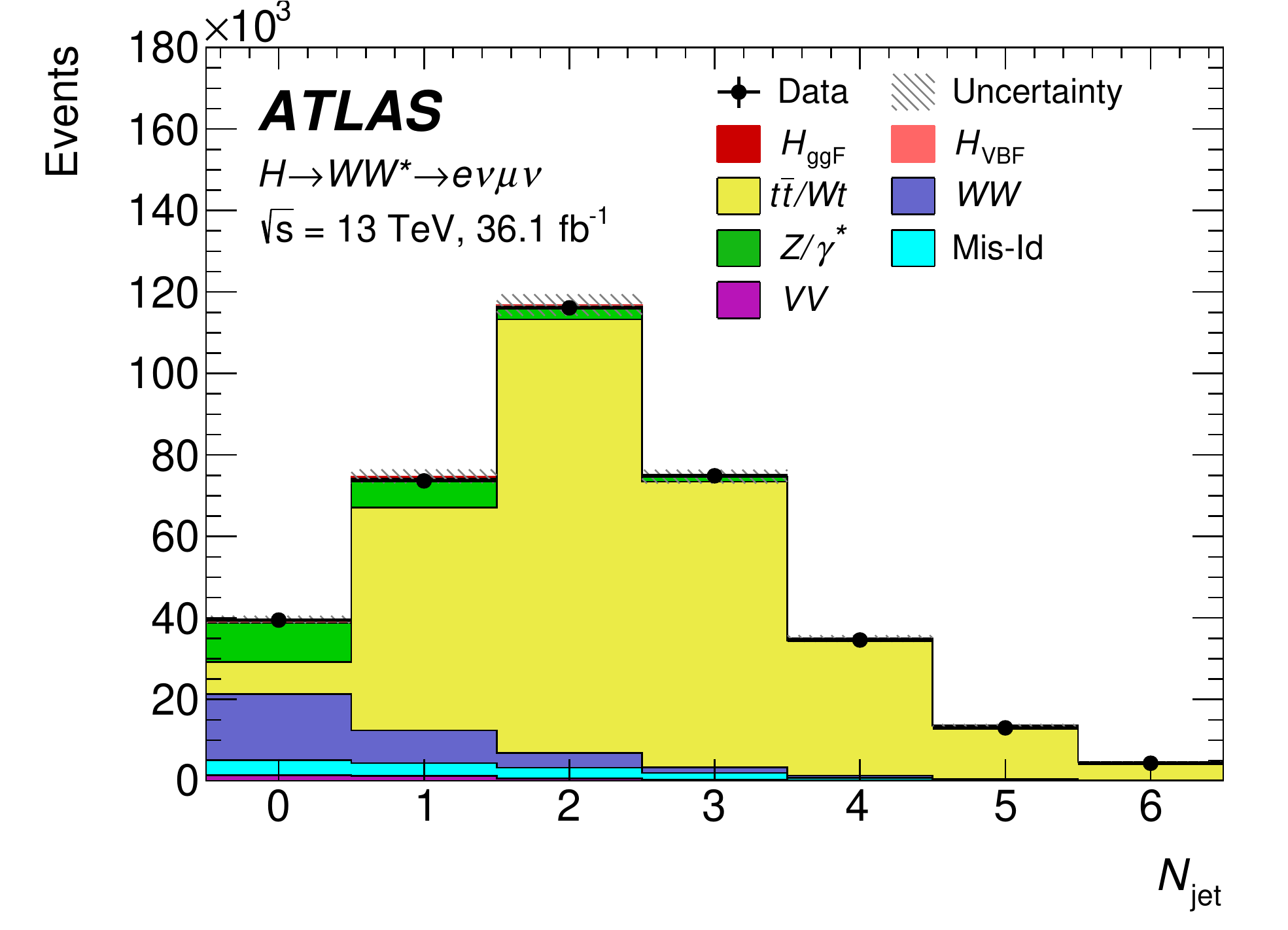}
{\caption{Jet multiplicity distribution after applying the preselection criteria. 
The shaded band represents the systematic uncertainty and accounts for experimental uncertainties only.
\label{fig:Plots:Njets}
 }}
\end{figure}
\begin{figure}[!h]
  \centering
  \subfloat[]{
    \includegraphics[width=0.5\textwidth]{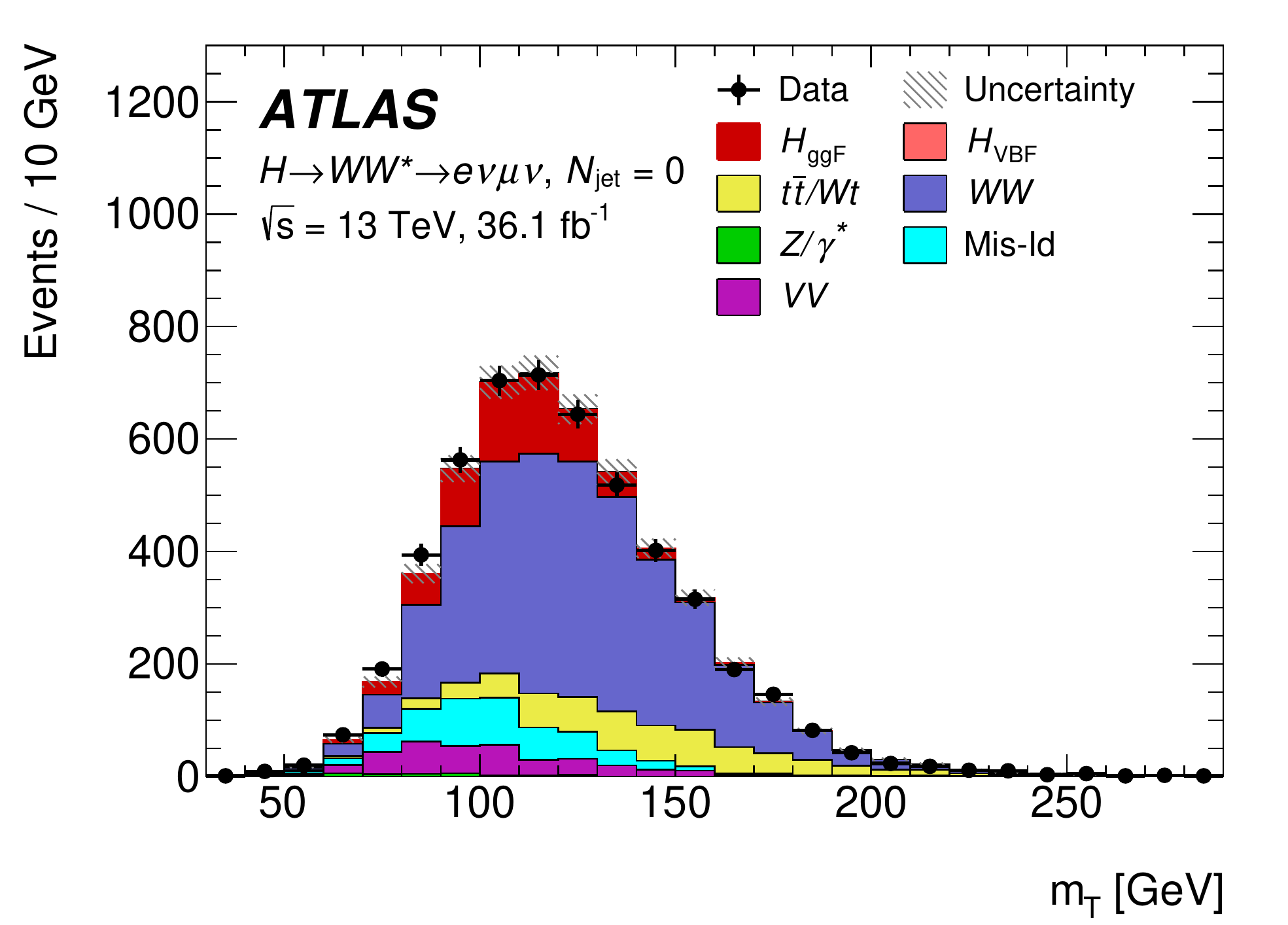}
  }
  \subfloat[]{
    \includegraphics[width=0.5\textwidth]{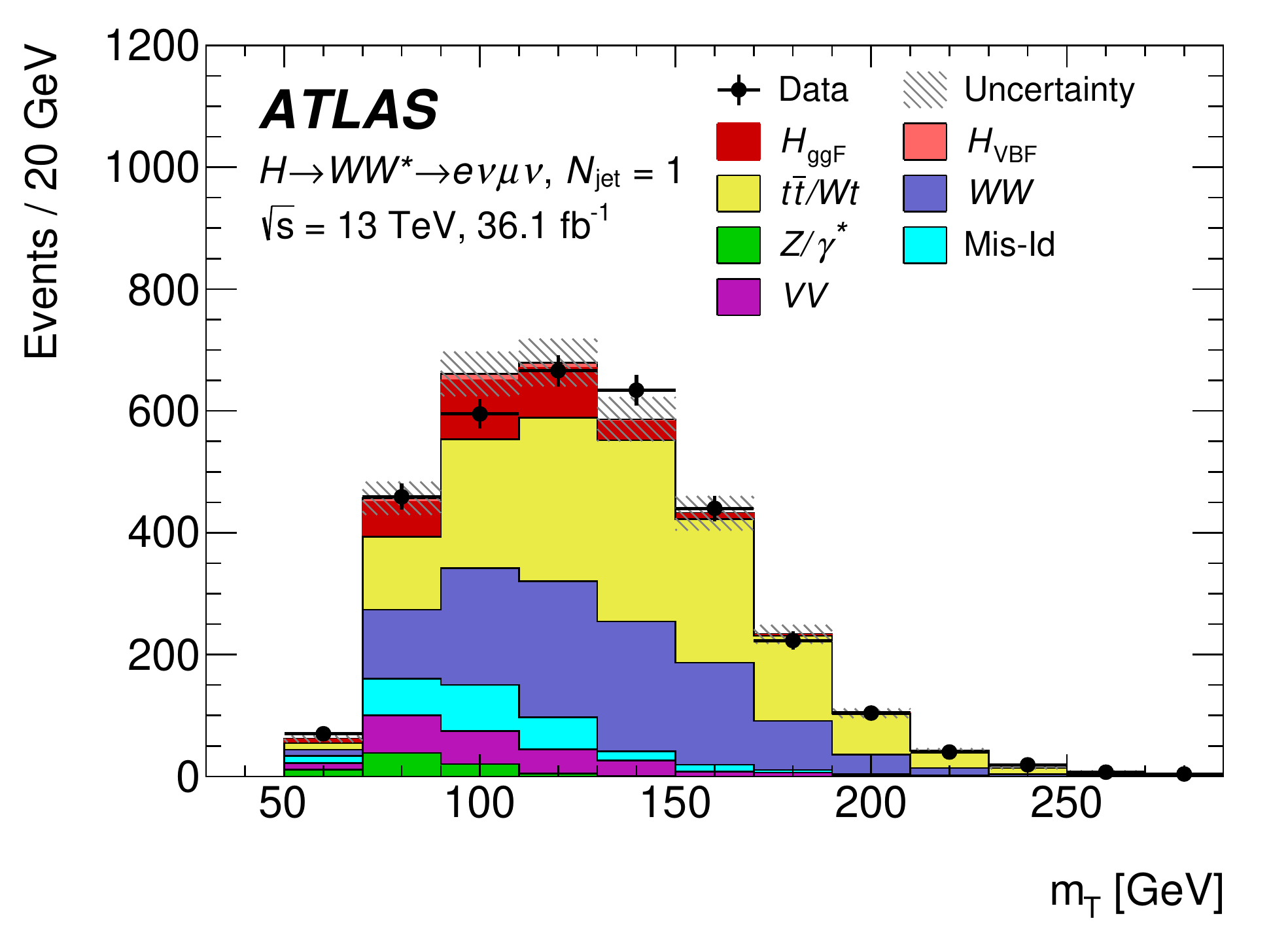}
  }
{\caption{Post-fit $\mT$ distributions with the signal and the background modelled contributions 
in the (a) $\ZeroJet$ and (b) $\OneJet$ signal regions.
The hatched band shows the total uncertainty of the signal and background modelled contributions. 
\label{fig:ggF_MT}
 }}
\end{figure}
\begin{figure}[!h]
  \centering
  \subfloat[]{
    \includegraphics[width=0.49\textwidth]{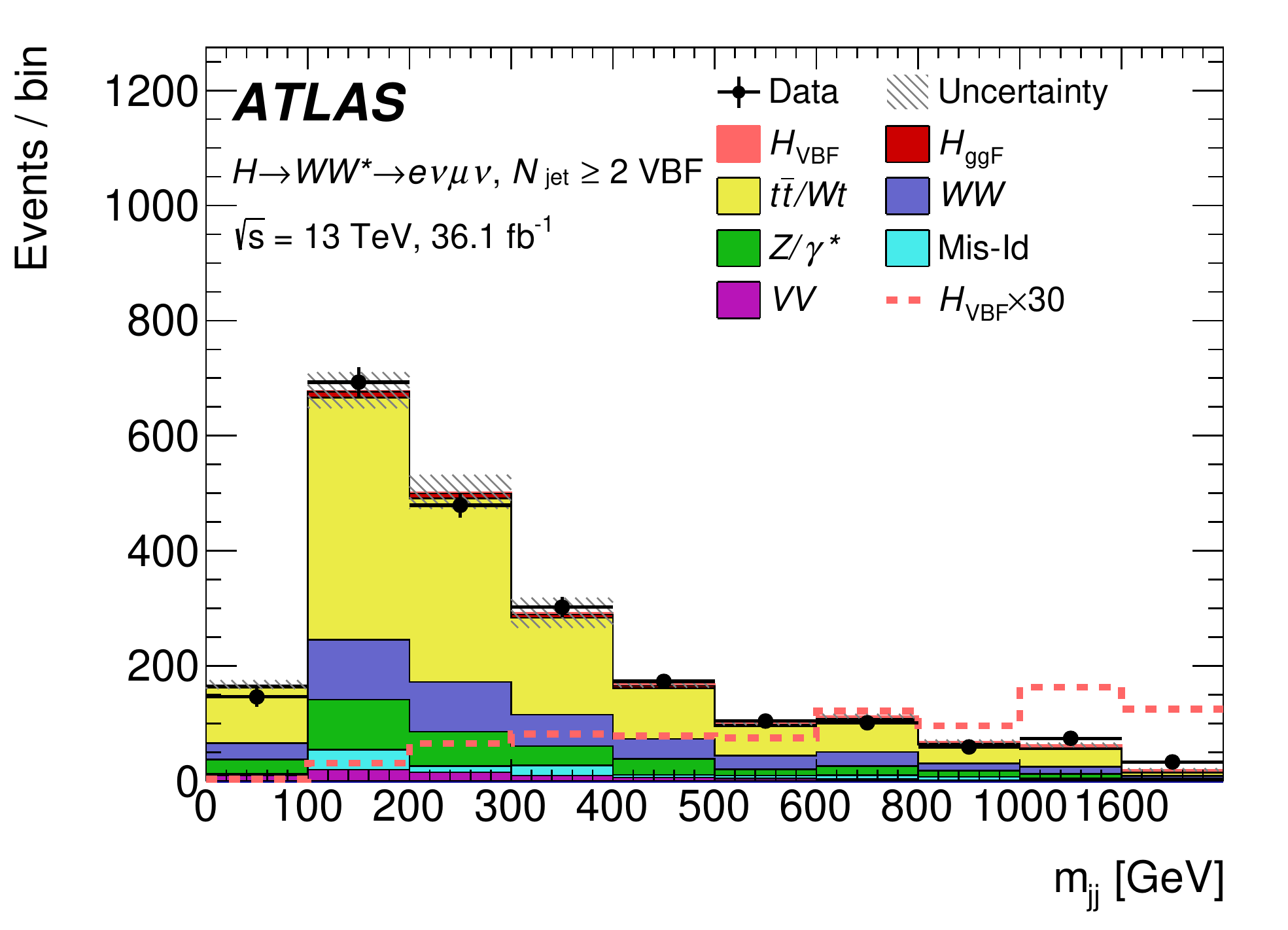}
  }
  \subfloat[]{
    \includegraphics[width=0.49\textwidth]{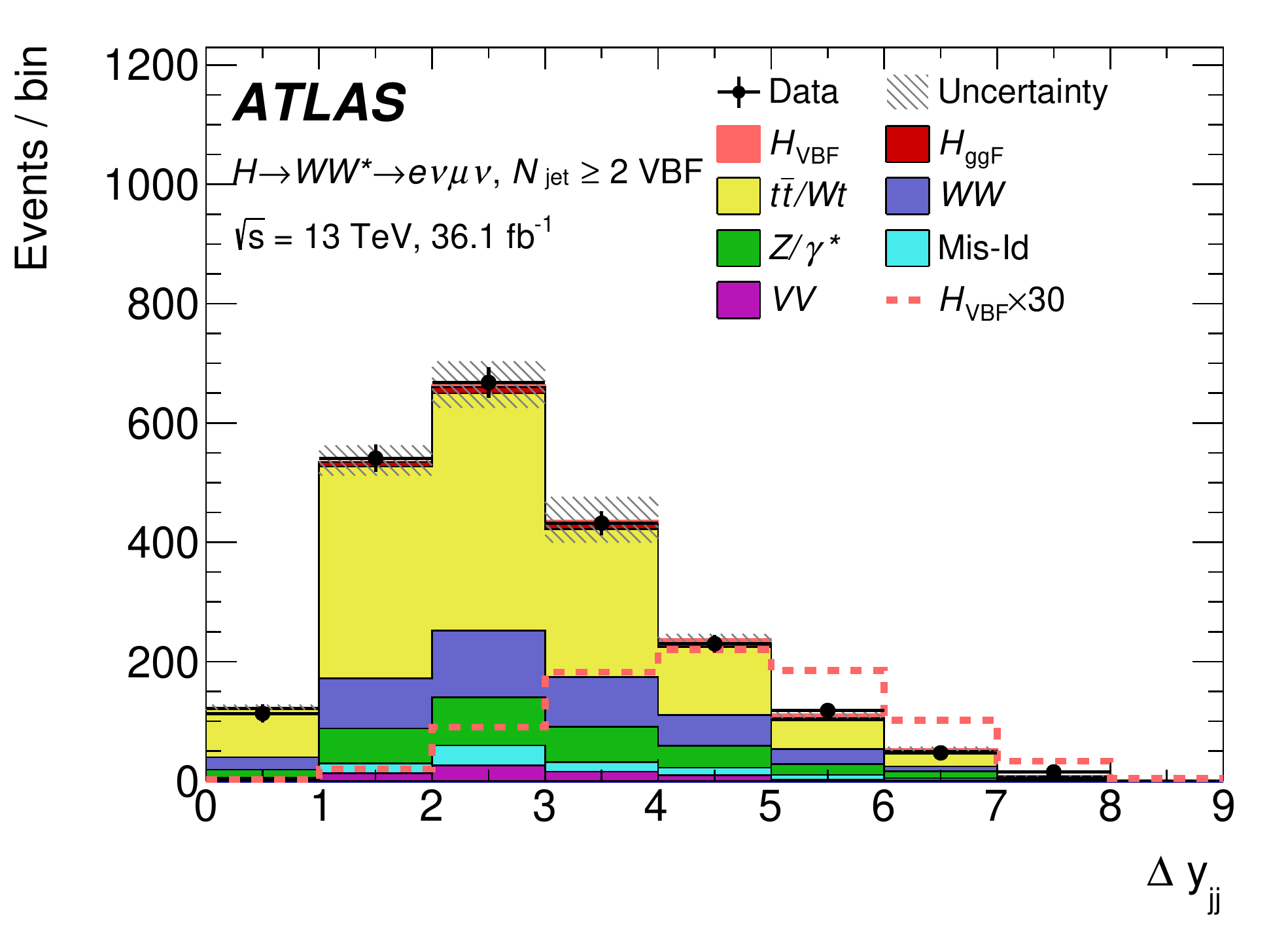}
  }
{\caption{
Post-fit $\mjj$ (a) and $\dyjj$ (b) distributions with signal and background modelled contributions
 in the $\TwoJet$ VBF signal region.
The dashed line shows the VBF signal scaled by a factor of 30.
The hatched band shows the total uncertainty of the signal and background modelled contributions.
\label{fig:vbf:Plots:SR}
 }}
\end{figure}
\begin{figure}[!h]
  \centering
  \includegraphics[width=0.55\textwidth]{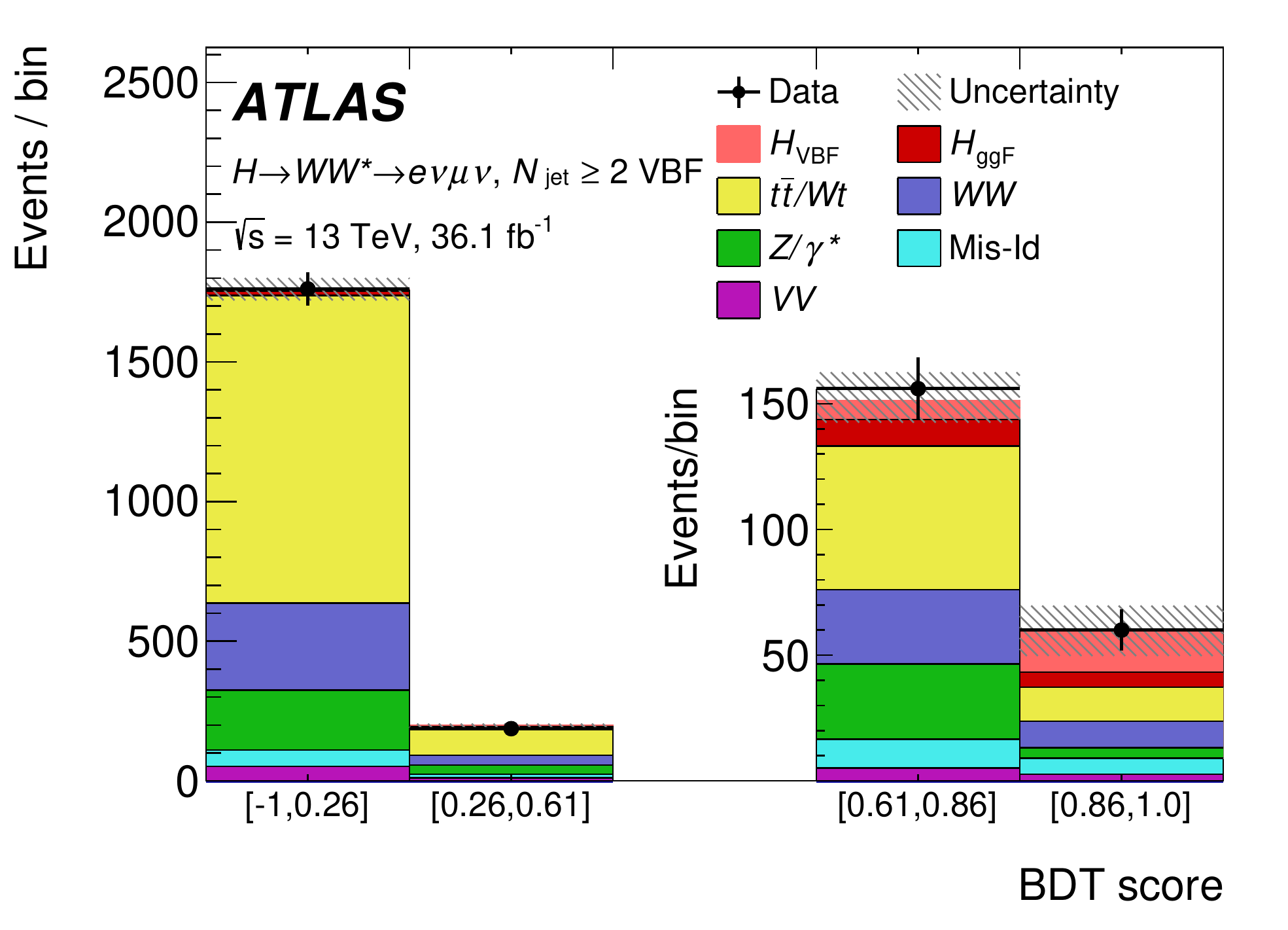}
  \caption{Post-fit BDT score distribution with the signal and the background modelled contributions 
in the VBF signal region.
The hatched band shows the total uncertainty of the signal and background modelled contributions. 
    \label{fig:VBF_BDT}
  }
\end{figure}

Events are triggered using single-lepton triggers and a dilepton $e$--$\mu$ trigger. 
The transverse momentum threshold ranges between $24~\GeV$ and $26~\GeV$ for single-electron triggers and between $20~\GeV$ and $26~\GeV$ for single-muon triggers, depending on the run period~\cite{Aaboud:2016leb}.
The $e$--$\mu$ trigger requires a minimum $\pt$ threshold of $17~\GeV$ for electrons and $14~\GeV$ for muons.\\  
Electron candidates are reconstructed from energy clusters in the electromagnetic calorimeter with an associated well-reconstructed track~\cite{PERF-2016-01,ATLAS-CONF-2016-024}. Electrons are required to satisfy $|\eta| < 2.47$, excluding the transition region between the barrel and endcap calorimeters, $1.37 < |\eta| < 1.52$.
Muon candidates are selected from tracks reconstructed in the ID matched to tracks reconstructed in the muon spectrometer~\cite{PERF-2015-10} and are required to satisfy $|\eta| < 2.5$. To reject particles misidentified as leptons, several identification requirements as well as calorimeter and track isolation criteria~\cite{ATL-PHYS-PUB-2015-041,PERF-2015-10} are applied. 
The electron identification criteria applied provide an efficiency in the range 88--94\% depending on electron $\pt$ and $\eta$. 
For muons, high efficiency, close to 95\%, is observed over the full instrumented $\eta$ range. 
The final lepton-selection criteria require two different-flavour opposite-sign leptons, the higher-$\pt$ (leading) lepton with $\pt > 22~\GeV$ and the subleading lepton with $\pt > 15~\GeV$. 
At least one of the leptons must correspond to a lepton that triggered the recording of the event. When the $e$--$\mu$ trigger is solely responsible for the recording of the event, each lepton must be matched to one of the trigger objects. 
The trigger matching requires the offline $\pt$ of the matching object to be higher than the trigger level threshold by at least $1~\GeV$. 
Jets are reconstructed using the anti-$k_t$ algorithm~\cite{Cacciari:2008gp} with a 
radius parameter $R = 0.4$.
The four-momenta of jets are corrected for the non-compensating response 
of calorimeter, signal losses due to noise threshold effects, energy lost in non-instrumented regions, and contributions from pile-up~\cite{PERF-2016-04}. 
Jets are required to have $\pt > 20~\GeV$ and $|\eta| <  4.5$.
A multivariate selection that reduces contamination from pile-up~\cite{Aaboud:2017pou} is applied to
jets with $\pt <  60~\GeV$ and $|\eta| <  2.4$, utilising calorimeter and tracking information to separate hard-scatter jets from pile-up jets. 
For jets with $\pt < 50~\GeV$ and $|\eta| >  2.5$, jet shapes and topological jet correlations in pile-up interactions are exploited to reduce contamination. 
Jets with $\pT > 20~\GeV$ and $|\eta| < 2.5$ containing $b$-hadrons ($b$-jets) are identified using a multivariate 
technique having as input the track impact parameters and information from secondary vertices. 
The adopted working point provides a nominal 3\% light-flavour ($u$-, $d$-, $s$-quark and gluon) misidentification rate and a 32\% $c$-jet misidentification rate with an average 85\% $b$-jet tagging efficiency, as estimated from simulated $t\bar{t}$ events~\cite{PERF-2016-05}. 
Ambiguities from overlapping reconstructed jet and lepton candidates are resolved as follows.
If a reconstructed muon shares an ID track with a reconstructed electron, the electron is removed. 
Reconstructed jets geometrically overlapping in a cone of radius $\Delta R = 0.2$ with electrons or muons are also removed. Electrons and muons, with transverse momentum $\pT$, are removed if they are within $\Delta R = \mathrm{min}(0.4, 0.04 + 10~\GeV/\pT)$ of the axis of any surviving jet. 
The missing transverse momentum $\vmet$ (with magnitude \met) is defined as the negative vector sum of the $\pt$ of all the selected leptons and jets, and including reconstructed tracks not associated with these objects, and consistent with originating from the primary $pp$ collision~\cite{Aaboud:2018tkc}. 
A second definition of missing transverse momentum (in this case denoted $\pTmiss$) uses the tracks associated with the jets instead of the calorimeter-measured jets. It was found during the optimisation that $\pTmiss$ performs better in terms of background rejection~\cite{Aaboud:2018tkc}.\\
Events are classified into one of three categories based on the number of jets with $\pt > 30~\GeV$: events with zero jets and events with exactly one jet target the ggF production mode ($\ZeroJet$ and $\OneJet$ ggF categories), and events with at least two jets target the VBF production mode ($\TwoJet$ VBF category). Figure~\ref{fig:Plots:Njets} shows the jet multiplicity distribution after applying the preselection criteria defined in Table~\ref{tab:HWWselection}. 
The different background compositions as a function of jet multiplicity motivate the division of the data sample into the various $N_\textrm{jet}$ categories and the definition of a signal region in each jet multiplicity bin. Details of the background estimation are provided in Section~\ref{sec:Background}.
To reject background from top-quark production, events containing $b$-jets with $\pt > 20~\GeV$ ($\Nbjetsub$) are vetoed. 
The full event selection is summarised in Table~\ref{tab:HWWselection},
where $\dphillmet$ is defined as the azimuthal angle between $\vmet$ and the dilepton system, $\pTll$ is the transverse momentum of the dilepton system, $\mll$ is the invariant mass of the two leptons, $\dphill$ is the azimuthal angle between the two leptons, and $\maxmtlep$ is the larger of $\mT^{\ell_i} = \sqrt{2\,\pt^{\ell_i}\cdot\met\cdot\left(1-\cos{\Delta\phi\left({\ell_i},\met\right)} \right)}$, where $\ell_i$ can be either the leading or the subleading lepton. 
The ``outside lepton veto'' requires the two leptons to reside within the rapidity gap spanned
by the two leading jets, and the ``central jet veto'' rejects events
with additional jets with $\pt > 20~\GeV$ in the rapidity gap between the two leading jets.
In the $\OneJet$ and $\TwoJet$ categories, the invariant mass of the $\tau$-lepton pair ($\mtt$), calculated using the collinear approximation~\cite{Plehn:1999xi}, is used to veto background from $\Ztt$ production. 
\begin{table}[!h]
\centering
\caption{
Event selection criteria used to define the signal regions in the \hwwenmn\ analysis.
For the $\TwoJet$ VBF signal region, the input variables used for the boosted decision tree (BDT) training are also reported.
\label{tab:HWWselection}
}
\scalebox{0.8}{
\input{figures/EvtSelection.tex}

}
\end{table}
Signal regions (SRs) are defined in each $\Njet$ category after applying all selection criteria. 
For both the $\ZeroJet$ and $\OneJet$ ggF SRs, eight regions, later used for the fit, are defined by subdividing in $\mll$ at $\mll < 30~\GeV$ and $\mll \geq 30~\GeV$, in $\pt$ of the subleading lepton at $\pTsublead < 20~\GeV$ and $\pTsublead \geq 20~\GeV$, and by the flavour of the subleading lepton. 
For the categories with zero jets and with exactly one jet, the discriminating variable between signal and SM background processes is the dilepton transverse mass, defined as $\mt = \sqrt{ {\left(E_\mathrm{{T}}^{\ell\ell} + E_\mathrm{{T}}^{\mathrm{miss}}\right)}^2 - {\left|{\mathbf p}_\mathrm{{T}}^{\ell\ell} + \vmet\right|}^2 }$ 
where $E_\mathrm{{T}}^{\ell\ell} = \sqrt{|{\mathbf p}_\mathrm{{T}}^{\ell\ell}|^2 + m_{\ell\ell}^2 }$ and ${\mathbf p}_\mathrm{{T}}^{\ell\ell}$ is the vector sum of the lepton transverse momenta. 
The discriminating variable $\mt$ is used in the ggF SRs, with eight bins for the $\ZeroJet$ and six bins for the $\OneJet$ regions. The bin boundaries are chosen such that approximately the same number of signal events is expected in each bin. 
The \mt distributions for the \ZeroJet and \OneJet SRs are shown in Figure~\ref{fig:ggF_MT}. All figures in this Letter, except Figure~\ref{fig:Plots:Njets}, use signal and background normalisations as fitted by the final statistical analysis of all signal and control regions, including pulls of statistical and systematic uncertainty parameters (post-fit).
For the $\TwoJet$ VBF selection, a boosted decision tree (BDT)~\cite{friedman2001} is used to enhance discrimination power between the VBF signal and backgrounds, including the ggF process. 
Kinematic variables of the two leading jets ($j$) and the two leading leptons ($\ell$) are used as inputs to the BDT: the invariant masses (\mjj, \mll), the difference between the two jet rapidities (\dyjj), and the difference between the azimuthal angles of the two leptons (\dphill).
Other variables used in the BDT training are: $\mt$,
the lepton $\eta$-centrality ($\lepetacent$, where $C_\ell = | 2\eta_\ell - \sum \eta_{j} | / \Delta \eta_{jj}$), which quantifies the positions of the leptons relative to the leading jets in pseudorapidity~\cite{BARGER1995106},
the sum of the invariant masses of all four possible lepton--jet pairs ($\summlj$), and 
the total transverse momentum (\pttot), which is defined as the magnitude of the vectorial sum of all selected objects. 
The observables providing the best discrimination between signal and background are 
$\mjj$ and $\dyjj$, and are shown in Figure~\ref{fig:vbf:Plots:SR} after applying all selections.
The BDT score reflects the compatibility of an event with VBF-like kinematics. 
Signal-like events would tend to have high BDT score, while background-like events tend to have low BDT score.
The signal purity, therefore, increases at high values of BDT score. 
The BDT score is used as the discriminating variable in the statistical analysis with four bins. 
The bin boundaries are chosen to maximise the expected sensitivity for the VBF production mode, resulting in smaller bin widths for larger values of the BDT score. 
In the highest-score BDT bin, the expected signal-to-background ratio of the VBF signal is approximately 0.6.  
The BDT distribution for the VBF-enriched region is presented in Figure~\ref{fig:VBF_BDT}.


%% file: figures/EvtSelection.tex
  \begin{tabular}{c|| c| c| c}
 \dbline
  Category          & $\NjetThirty=0$ ggF & $\NjetThirty=1$ ggF & $\NjetThirty\ge2$ VBF \\

  \sgline
  \multirow{2}{*}{ Preselection}     &
  \multicolumn{3}{c}{
  \begin{tabular}{c}
  Two isolated, different-flavour leptons ($\ell\,{=}\,e, \mu$) with opposite charge\\
  $\pTlead\,{>}\,22~\GeV$ , $\pTsublead\,{>}\,15~\GeV$ \\
  $\mll\,{>}\,10~\GeV$ \\
  \end{tabular}
  }\\
  & \multicolumn{2}{c|}{$\pTmiss\,{>}\,20~\GeV$}  & \\
 \sgline
  \multirow{3}{*}{Background rejection}  & \multicolumn{3}{c}{$\Nbjetsub=0$}       \\
                        & $\dphillmet > \pi/2$  &  $\maxmtlep\,{>}\,50~\GeV$         & \\
 	                & $\pTll\,{>}\,30~\GeV$                 &  \multicolumn{2}{c}{\hspace{-3.5cm}$\mtt\,{<}\,m_Z - 25~\GeV$}    \\
  \sgline
  \multirow{2}{*}{\!\!\!\!\!
    \begin{tabular}{c}
   $\hwwenmn$  \\topology
\end{tabular}
}                          & \multicolumn{2}{c|}{$\mll\,{<}\,55~\GeV$}    &  central jet veto\\
                           & \multicolumn{2}{c|}{$\dphill\,{<}\,1.8$}    & outside lepton veto\\
 \sgline
Discriminant variable   & \multicolumn{2}{c|}{$\mt$}    & BDT  \\
BDT input variables     &  \multicolumn{2}{c|}{ }       & $\mjj$, $\dyjj$, $\mll$, $\dphill$, $\mt$, $\lepetacent$, $\summlj$, $\pttot$ \\
  \dbline
  \end{tabular}

%% file: sections/Background_Letter.tex
\begin{figure}[H]
  \centering
  \subfloat[]{
    \includegraphics[width=0.5\textwidth]{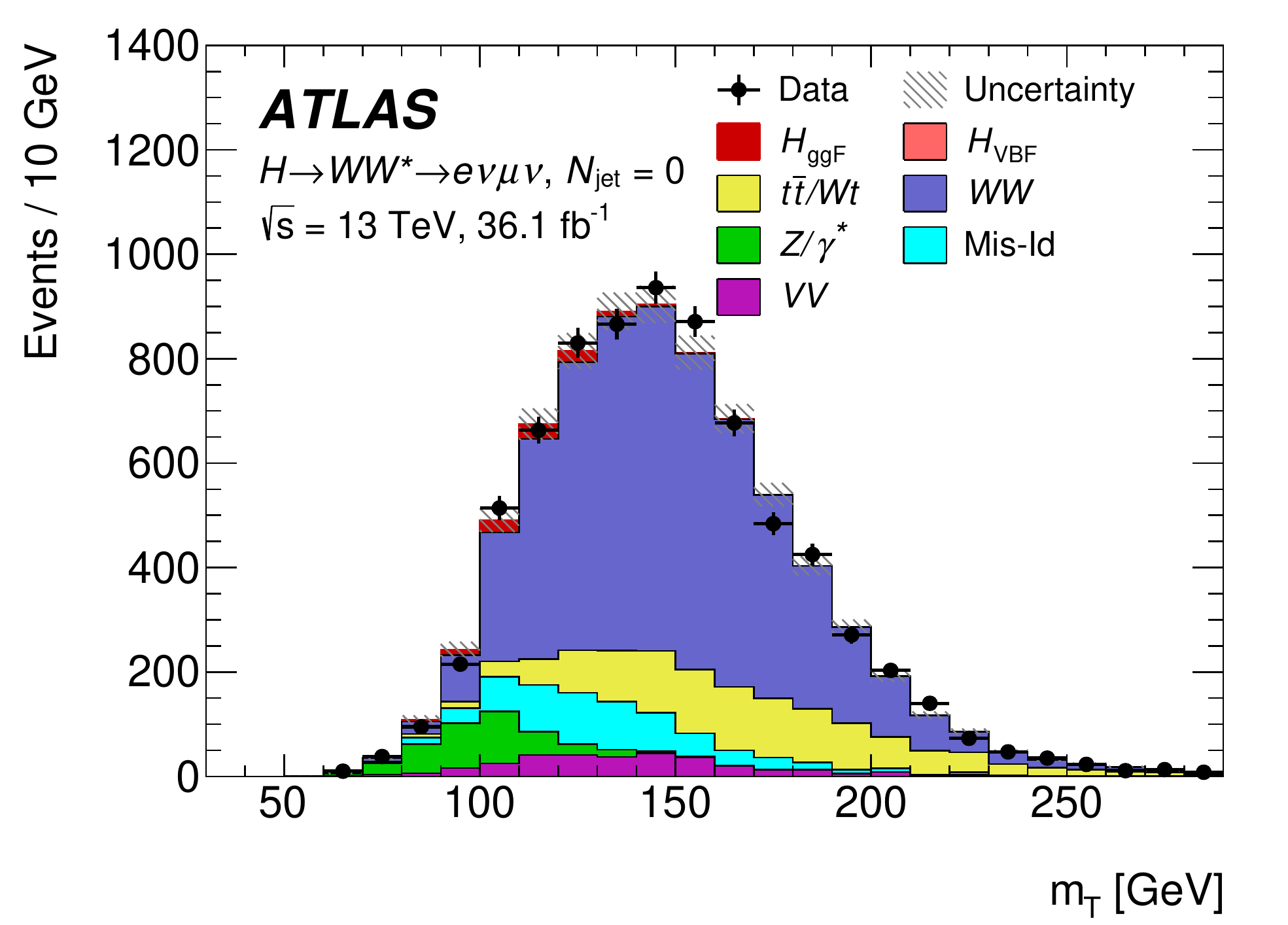}
  }
  \subfloat[]{
    \includegraphics[width=0.5\textwidth]{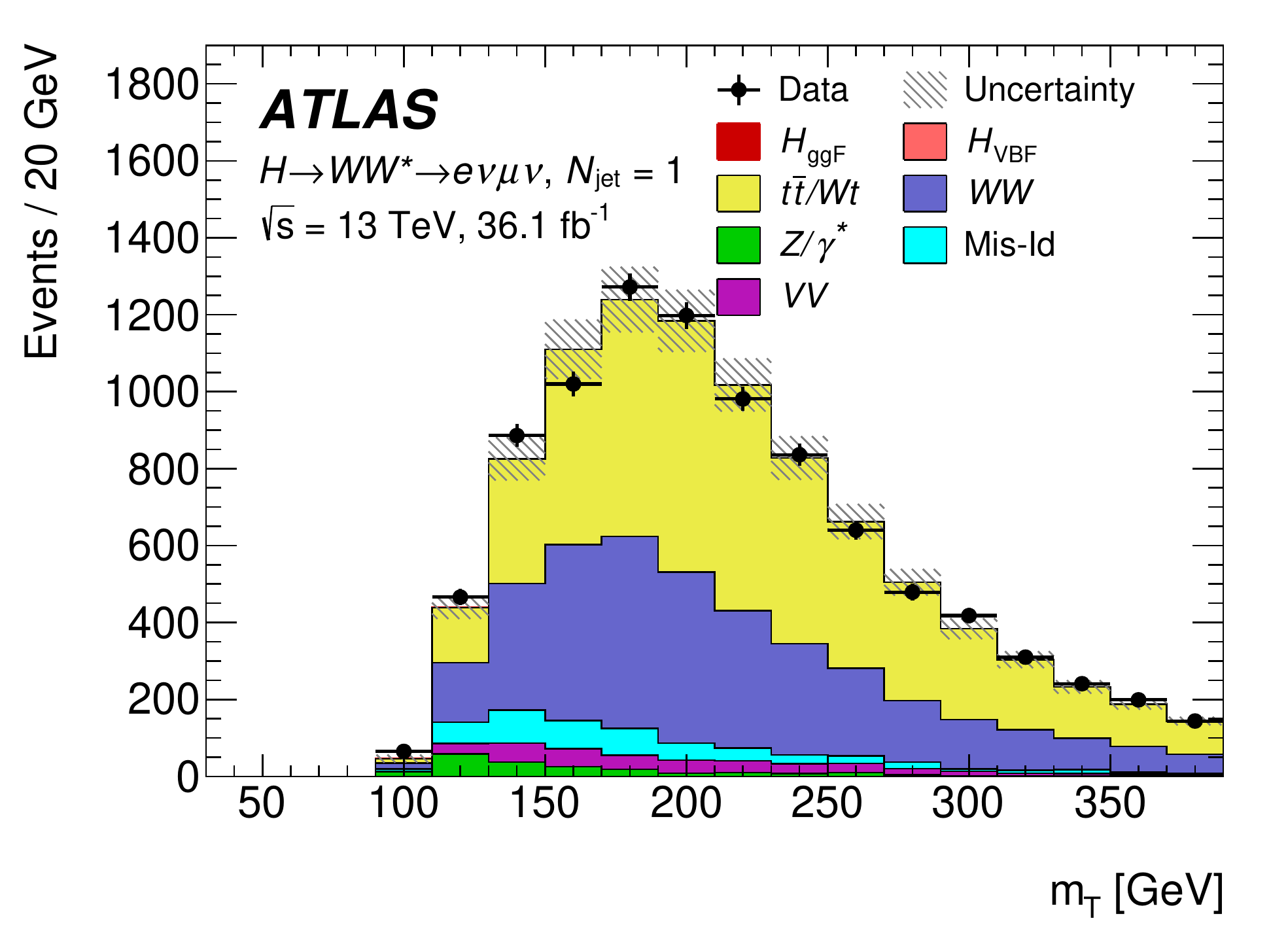}
  } \\
  \subfloat[]{
    \includegraphics[width=0.5\textwidth]{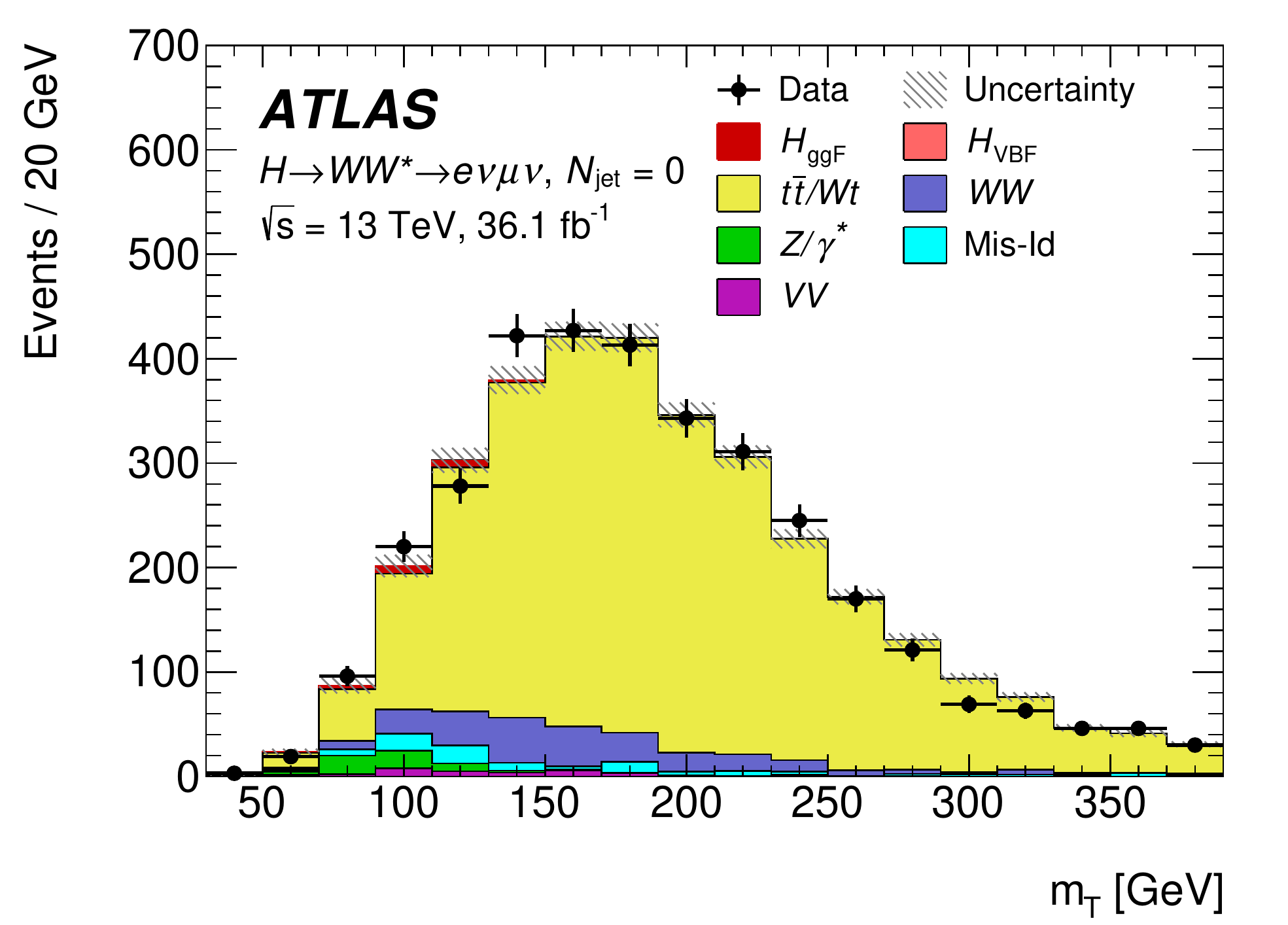}
  }
  \subfloat[]{
    \includegraphics[width=0.5\textwidth]{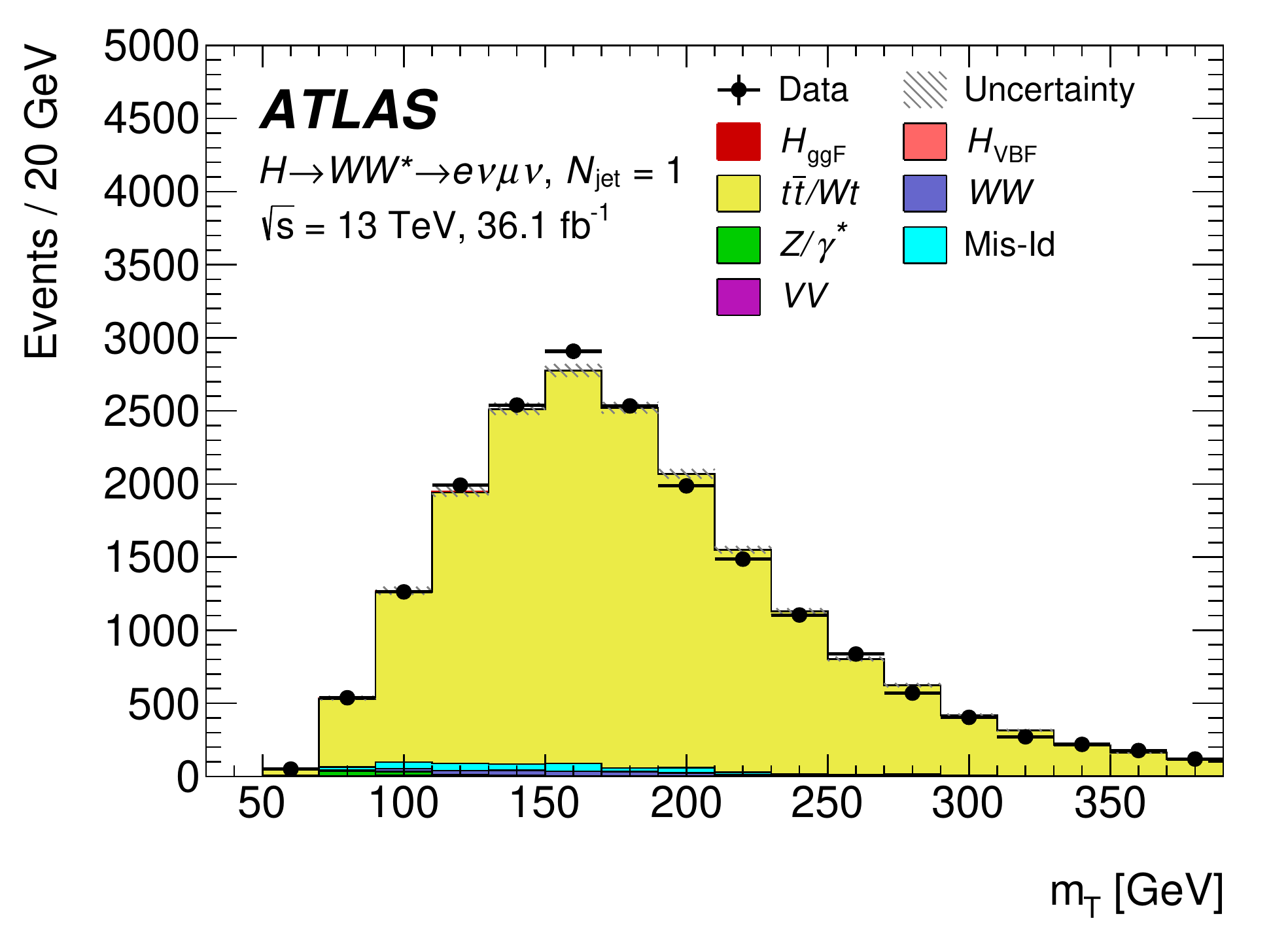}
  } \\
  \subfloat[]{
    \includegraphics[width=0.5\textwidth]{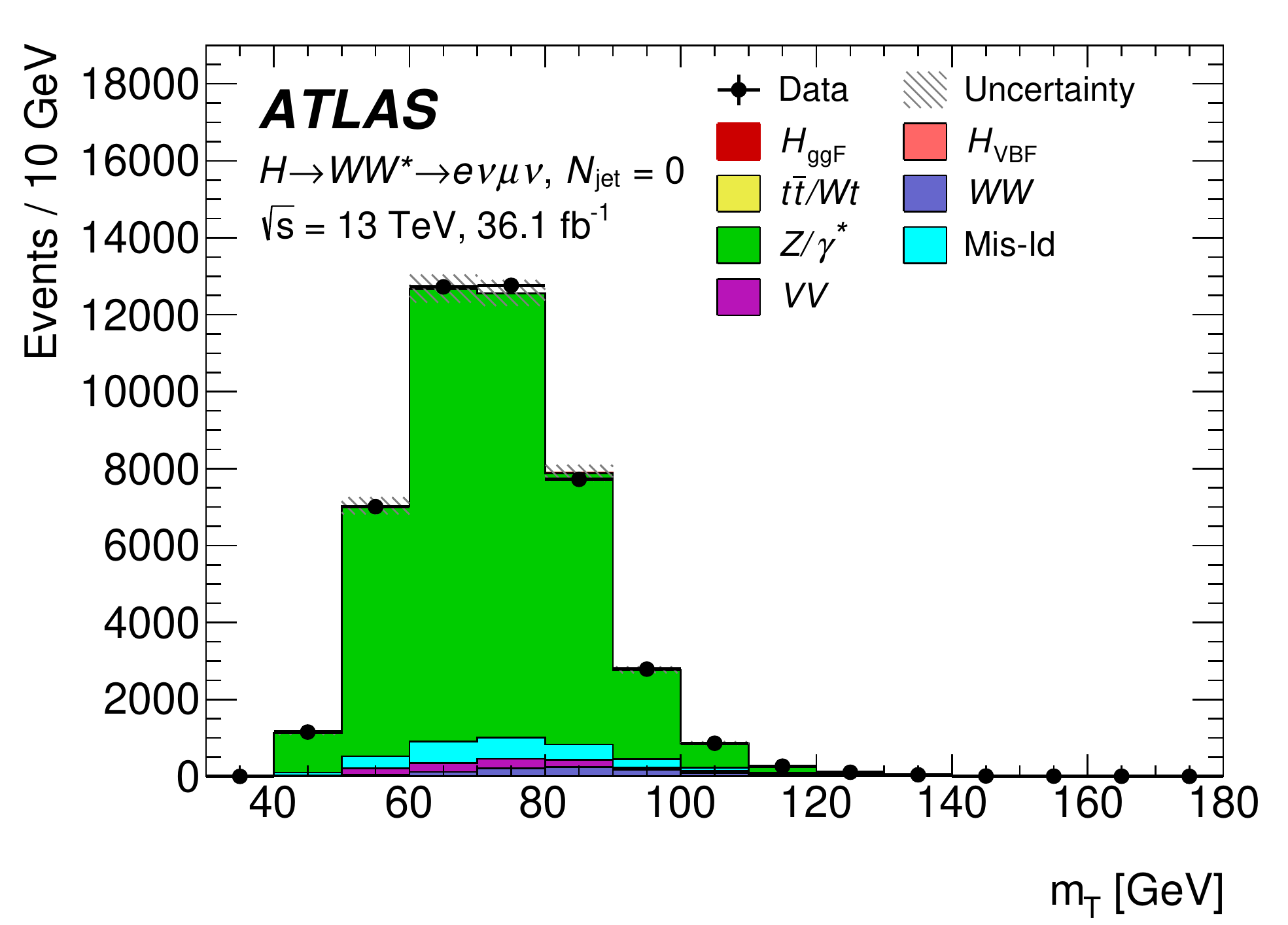}
  }
  \subfloat[]{
    \includegraphics[width=0.5\textwidth]{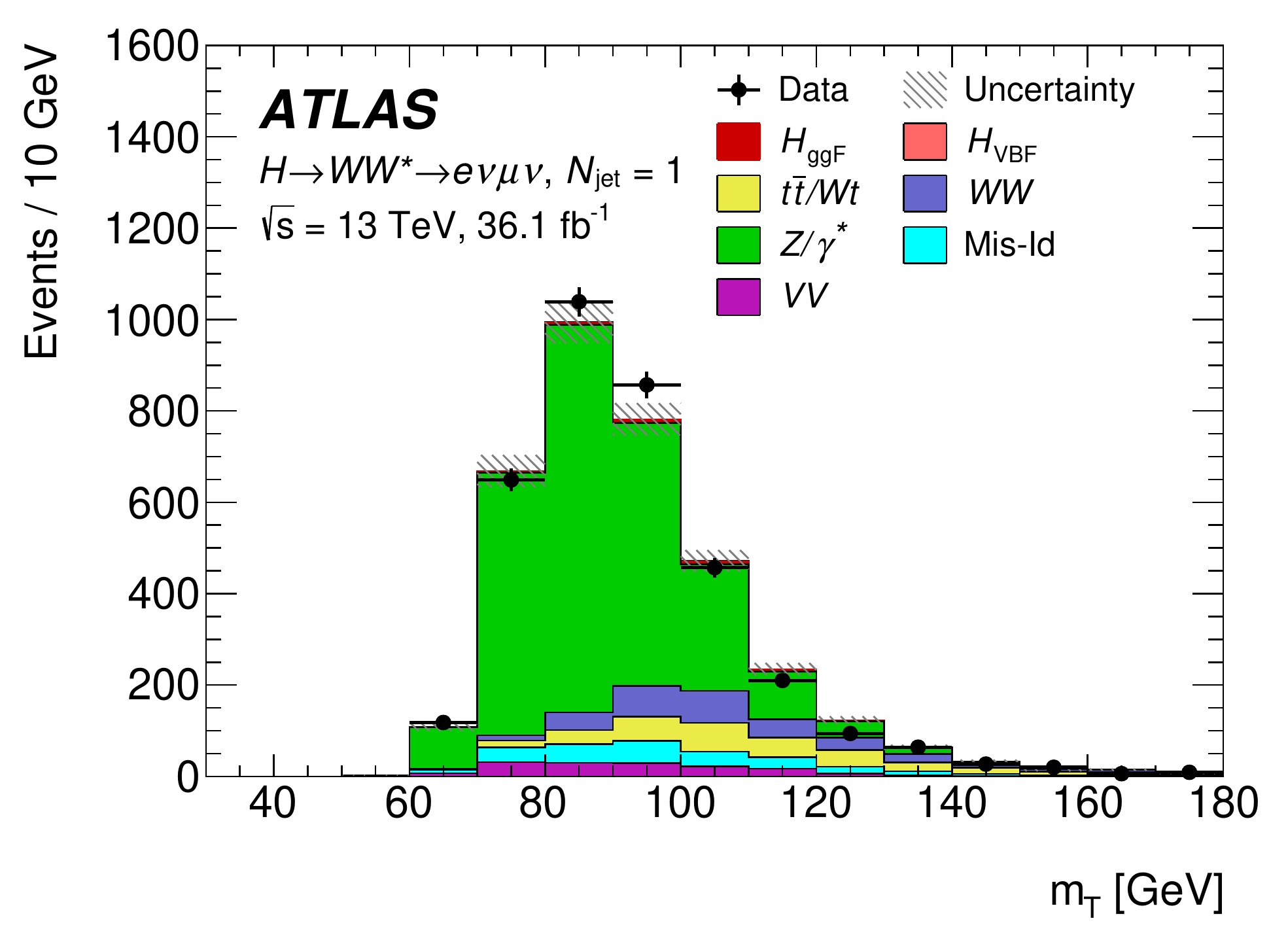}
  }
{\caption{Post-fit $\mT$ distributions with signal and and background modelled contributions in the $\ZeroJet$ and $\OneJet$ control regions for the $WW$ (a,b), $t\bar{t}$/$Wt$ (c,d), and $Z/\gamma^{\ast}$ (e,f) processes. 
The hatched band shows the total uncertainty of the signal and background modelled contributions. 
Some contributions are too small to be visible.
\label{fig:ggf:Plots:CR}
 }}
\end{figure}
\begin{figure}[!h]
  \centering
  \subfloat[]{
    \includegraphics[width=0.5\textwidth]{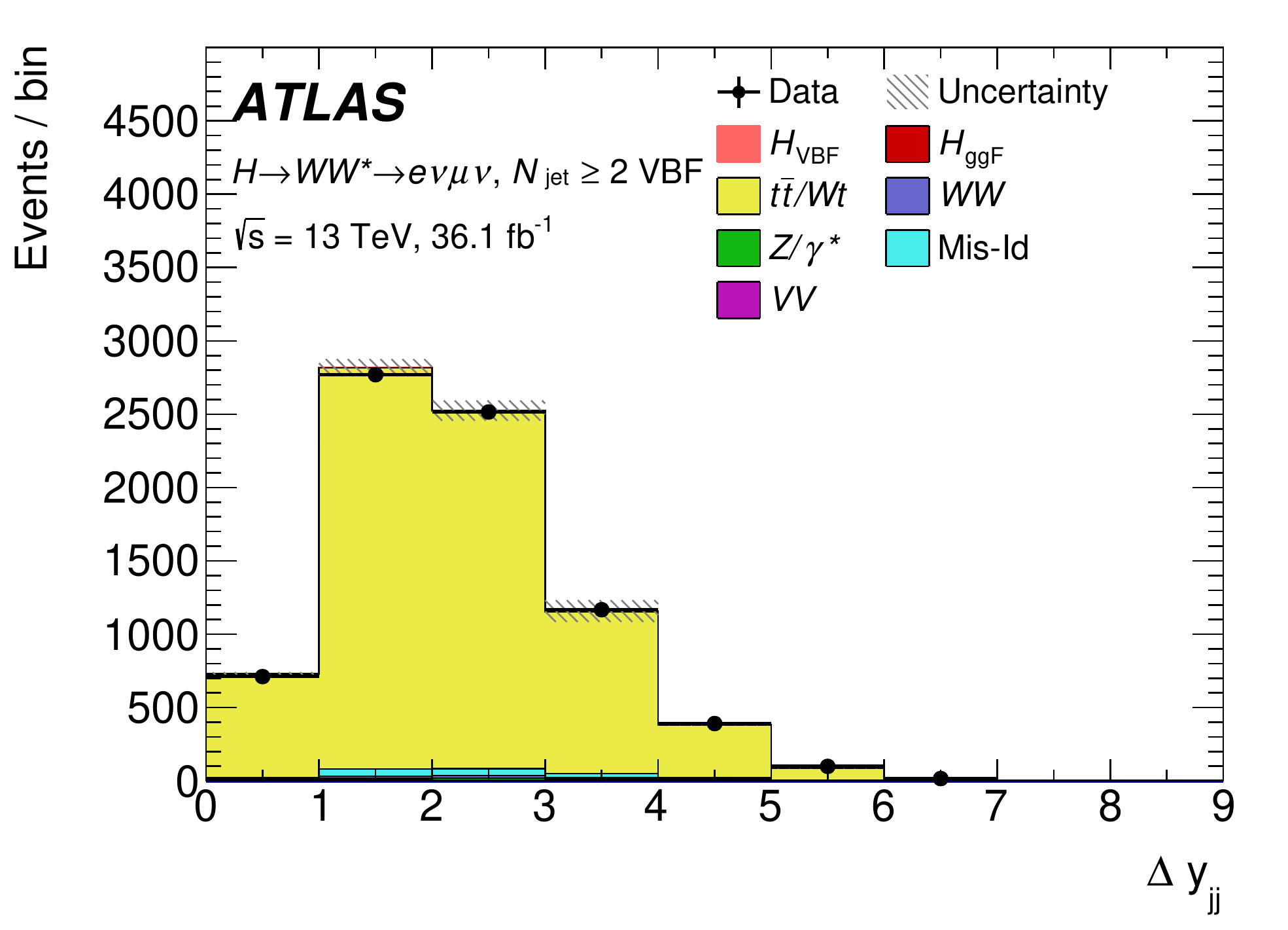}
  }
  \subfloat[]{
    \includegraphics[width=0.5\textwidth]{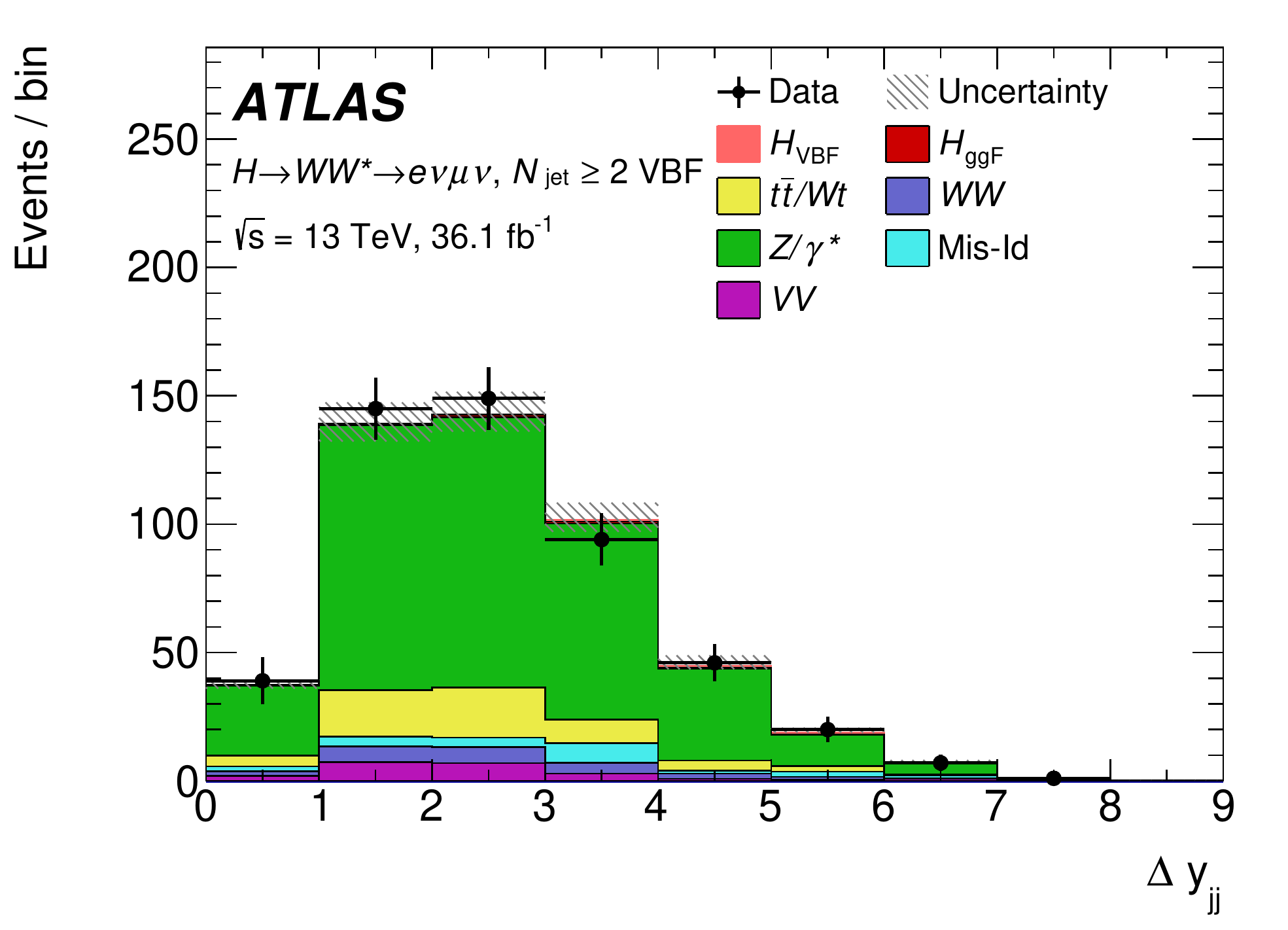}
  }
\caption{
Post-fit $\dyjj$ distribution with signal and background modelled contributions 
in the (a) $t\bar{t}$/$Wt$ and (b) $Z/\gamma^{\ast}$ control regions in the $\TwoJet$ VBF analysis category.
The hatched band shows the total uncertainty of the signal and background modelled contributions. 
Some contributions are too small to be visible. 
\label{fig:vbf:Plots:TopZCR}
 }
\end{figure}

The background contamination in the SRs originates from various processes: non-resonant $WW$,
top-quark pair ($t\bar{t}$) and single-top-quark ($Wt$), diboson ($WZ$, $ZZ$, $W\gamma$ and $W\gamma^{\ast}$) and Drell--Yan (mainly $\Ztt$, hereafter denoted $Z/\gamma^{\ast}$) production.
Other background contributions arise from $W+$jets and multi-jet production with misidentified leptons, 
which are either non-prompt leptons from decays of heavy-flavour hadrons or jets faking prompt leptons. 
\begin{table}[!h]
  \caption{
    Event selection criteria used to define the control regions.
    Every control region selection starts from the selection labelled ``Preselection'' in Table~\ref{tab:HWWselection}. 
    $\Nbjetbetween$ represents the number of $b$-jets with $20~\GeV< \pT < 30~\GeV$.
  }
  \label{tab:CRsSelection}
\centering
\scalebox{0.8}{
\input{figures/EvtSelectionCRs.tex}

}
\end{table}
Dedicated regions in data, identified hereafter as control regions (CRs), are used to normalise the predictions of some of the background processes. 
CRs are defined for the main background processes: $WW$ (only for $\ZeroOneJetSimple$ final states), $t\bar{t}$/$Wt$, and $Z/\gamma^{\ast}$. 
Table~\ref{tab:CRsSelection} summarises the event selection for all CRs.
For the $\ZeroJet$ and $\OneJet$ $WW$ CRs, \mll selections orthogonal to those of the SRs are applied.
For the $t\bar{t}$/$Wt$ CRs, the $b$-veto is replaced with a $b$-tag requirement.
For the $\OneJet$ and $\TwoJet$ VBF $Z/\gamma^{\ast}$ CRs, the $\mtt$ selection is inverted, while for the $\ZeroJet$ $Z/\gamma^{\ast}$ CR the $\dphill$ selection criterion is inverted. Figure~\ref{fig:ggf:Plots:CR} presents the post-fit $\mT$ distributions in the $\ZeroJet$ and $\OneJet$ CRs. \\
In Figure~\ref{fig:vbf:Plots:TopZCR}, the post-fit $\dyjj$ distributions in the $\TwoJet$ VBF CRs are shown. 
Data and simulation are in agreement within uncertainties for all the relevant distributions in the different CRs.
The background contributions with misidentified leptons are estimated using a data-driven technique. A control sample where one of the two lepton 
candidates fails to meet the nominal identification and isolation criteria but satisfies looser identification criteria, referred as an anti-identified lepton, is used. The contribution of this background in the SRs and CRs is then obtained by scaling the number of data events, after the subtraction of processes with two prompt leptons, in the control samples by an extrapolation factor. 
The latter is measured in a $\Zjets$-enriched data sample, where the $Z$ boson decays to a pair of electrons or muons, and the misidentified lepton candidate recoils against the $Z$ boson. 
The extrapolation factor is defined as the ratio of the numbers of identified 
and anti-identified leptons, and is measured in bins of $\pt$ and $\eta$. 
Furthermore, a sample composition correction factor is applied separately in $\pt<25~\gev$ and $\pt>25~\gev$ bins, and is defined in each bin as the ratio of the extrapolation factors measured in $\Wjets$ and $\Zjets$ MC simulation. 
The total uncertainty of the background with misidentified leptons includes uncertainties due to the difference in sample composition between the $W$+jets and $Z$+jets control samples determined with MC simulation, the statistical uncertainty of the $Z$+jets control sample, and 
the subtraction of other processes. 
In the VBF regions, the background estimation is corrected for the contamination from events with two misidentified leptons, whose origin is largely multi-jet events. This contribution is negligible in other regions. 
Details of this method can be found in Ref.~\cite{HIGG-2013-13}. \\
\begin{table}[!h]
\centering 
\caption{Post-fit normalisation factors which scale the corresponding estimated yields in the signal region; the dash indicates where MC-based normalisation is used. 
 The errors include the statistical and systematic uncertainties.}
\label{tab:CRs_NF}
 \scalebox{0.8}{
\begin{tabular}{c c c c}
\hline\hline
Category      & $WW$           &   $t\bar{t}/Wt$   & $Z/\gamma^{\ast}$   \\  
\hline\hline
$\NjetThirty=0$ ggF    &  1.06 $\pm$ 0.09 &  0.99 $\pm$ 0.17 & 0.84 $\pm$ 0.04 \\
$\NjetThirty=1$ ggF    &  0.97 $\pm$ 0.17 &  0.98 $\pm$ 0.08 & 0.90 $\pm$ 0.12 \\
$\NjetThirty\ge2$ VBF  &  --              &  1.01 $\pm$ 0.01 & 0.93 $\pm$ 0.07  \\
\hline\hline
\end{tabular}}
\end{table}
The post-fit background normalisation factors are summarised in Table~\ref{tab:CRs_NF}.  
The $Z/\gamma^{\ast}$ normalisation factors are affected by residual misalignments in the inner detector which distort the measurements of the track parameters for particles originating from secondary vertices e.g. leptons from $\tau$ decays.

%% file: figures/EvtSelectionCRs.tex
  \begin{tabular}{c|| c| c |c}
  \dbline
  CR                         & $\NjetThirty=0$ ggF & $\NjetThirty=1$ ggF & $\NjetThirty\ge2$ VBF \\
  \sgline
  \multirow{4}{*}{$WW$}      & $55\,{<}\,\mll\,{<}\,110~\GeV$  & $\mll\,{>}\,80~\GeV$        &   \\
                             & $\dphill\,{<}\,2.6$             & $|\mtt-m_Z|\,{>}\,25~\GeV$  &   \\
                             & \multicolumn{2}{c|}{$\Nbjetsub=0$}            &   \\
                             &                                 & $\maxmtlep\,{>}\,50~\GeV$  &   \\
  \sgline
  \multirow{5}{*}{$t\bar{t}$/$Wt$} & \multirow{2}{*}{$\Nbjetbetween>0$}    & $\Nbjetnom = 1$                           & \multirow{2}{*}{$\Nbjetsub=1$}        \\
                             &                                       & $\Nbjetbetween  = 0$                      &                      \\
                             & $\dphillmet > \pi/2$                  & $\maxmtlep\,{>}\,50~\GeV$                 &  central jet veto    \\
                             & $\ptll\, > \, 30~\GeV$                & \multicolumn{2}{c}{\phantom{ccccc}$\mtt\,{<}\,m_Z - 25~\GeV$}                    \\
                             & $\dphill\,{<}\,2.8$                   &                                           &  outside lepton veto  \\
  \sgline
  \multirow{4}{*}{$Z/\gamma^{\ast}$}    &                        \multicolumn{3}{c}{\phantom{ccccc}$\Nbjetsub=0$}           \\
                             & \multicolumn{3}{c}{\phantom{ccccc}$\mll < 80~\GeV$}             \\
                             & \multicolumn{2}{c}{no $\pTmiss$ requirement}                     &  {central jet veto} \\
                             &                                  &  $\maxmtlep\,{>}\,50~\GeV$   &  \multicolumn{1}{c}{outside lepton veto} \\
                             & $\dphill > 2.8$        & $\mtt > m_Z - 25~\GeV$ &$|\mtt - m_Z| \leq 25~\GeV$  \\

\dbline
  \end{tabular}

%% file: sections/Systematics_Letter.tex
The sources of uncertainty can be classified into two categories: experimental and theoretical.
The dominant experimental uncertainties are the jet energy scale and resolution~\cite{ATL-PHYS-PUB-2015-015}, and the $b$-tagging efficiency~\cite{ATL-PHYS-PUB-2016-012}.
Other sources of uncertainty are lepton energy (momentum) scale and resolution, identification and isolation~\cite{ATL-PHYS-PUB-2016-015,ATLAS-CONF-2016-024,PERF-2015-10}, missing transverse momentum measurement~\cite{ATL-PHYS-PUB-2015-027}, modelling of pile-up, and luminosity measurement~\cite{Aaboud:2016hhf}.
The luminosity uncertainty is only applied to the Higgs boson signal and to background processes that are normalised to theoretical predictions.
For the main processes, the theoretical uncertainties are assessed by a comparison between nominal and alternative event generators and UEPS models, as indicated in Table~\ref{tab:mcsamples}. 
For the prediction of $WZ$, $ZZ$, $V\gamma^{\ast}$, and $V\gamma$ production ($VV$), variations of the matching scale are considered instead of an alternative generator.
In addition, the effects of QCD factorisation and renormalisation scale variations and PDF model uncertainties are evaluated.

%% file: sections/SignalExtraction_Letter.tex
The ggF and VBF cross-sections are obtained from a simultaneous statistical analysis of the data samples in all SRs and CRs by maximising a likelihood function in a fit using scaling parameters multiplying the predicted total production cross-section of each signal process and applying the profile likelihood method. 
The CRs are used to determine the normalisation of the corresponding backgrounds. 
The systematic uncertainties enter the fit as nuisance parameters in the likelihood function.\\ 
Table~\ref{tab:ggF_VBF_yields} shows the post-fit yields for all of the three SRs. Yields in the highest-score VBF BDT bin are also given.  
The uncertainties in the total yields are smaller than those of some of the individual background processes. 
This effect is due to correlations among different data regions, background processes, and nuisance parameters. 
The correlations are imposed by the fit as it constrains the total yield to match the data.
 For example, for the $b$-tagging efficiency, which is the main source of uncertainty in the $t\bar{t}/Wt$ yields in the SRs as well as in $WW$ CRs, 
the combination of these two regions in the statistical analysis leads to an anti-correlation between the SR yields of the $WW$ and $t\bar{t}/Wt$ backgrounds. 
Changes in the $b$-tagging efficiency simultaneously increase/decrease the yields of $t\bar{t}/Wt$ and $WW$ backgrounds, resulting in a small uncertainty in the combined yields of the processes but large uncertainties in the individual components.
\begin{table}[t!]
  \centering
  \caption{ 
Post-fit MC and data yields in the ggF and VBF SRs. Yields in the highest-score VBF BDT bin are also presented. 
The quoted uncertainties include the theoretical and experimental systematic sources and those due to sample statistics.
The sum of all the contributions may differ from the total value due to rounding. 
Moreover, the total uncertainty differs from the sum in quadrature of the single-process uncertainties due to the correlations. 
 \label{tab:ggF_VBF_yields}}
 \scalebox{0.8}{
 \begin{tabular}{lS[table-format=4,table-number-alignment=right]@{$\,\pm\,$}
                     S[table-format=3,table-number-alignment=left]
                     S[table-format=4,table-number-alignment=right]@{$\,\pm\,$}
                     S[table-format=3,table-number-alignment=left]
                     S[table-format=4,table-number-alignment=right]@{$\,\pm\,$}
                     S[table-format=2,table-number-alignment=left]
                     S[table-format=2,table-number-alignment=right]@{$\,\pm\,$}
                     S[table-format=2,table-number-alignment=left]}
  \dbline
  Process & \multicolumn{2}{c}{\ZeroJet ggF} & \multicolumn{2}{c}{\OneJet ggF} & \multicolumn{4}{c}{\TwoJet VBF}  \tabularnewline
          & \multicolumn{2}{c}{}  &  \multicolumn{2}{c}{}         &  \multicolumn{2}{c}{Inclusive} &\multicolumn{2}{c}{BDT: $[0.86,1.0]$}\tabularnewline
  \sgline
  $H_{\mathrm{ggF}}$  &    639 & 110             &  285 & 51                 &  42   & 16 & 6   &  3   \tabularnewline
  $H_{\mathrm{VBF}}$  &      7 & 1               &   31 & 2                  &  28   & 16 & 16  &  6   \tabularnewline
  \sgline                                               
  $WW$           &   3016 & 203             & 1053 & 206                     &  400  & 60 & 11  &  2   \tabularnewline
  $VV$           &   333  & 38              &  208 & 32                      &  70   & 12 & 3   &  1   \tabularnewline
  $t\bar{t}/Wt$  &   588  & 130             & 1397 & 179                     &  1270 & 80 & 14  &  2   \tabularnewline
  Mis-Id         &   447  & 77              &  234 & 49                      &  90   & 30 & 6   &  2   \tabularnewline
  $Z/\gamma^{*}$ &   27   & 11              &  76  & 24                      &  280  & 40 & 4   & 1   \tabularnewline
  \sgline
  Total          &   5067 & 80              & 3296 &  61                     &  2170 & 50 & 60  &  10  \tabularnewline
  Observed  &  \multicolumn{2}{l}{5089}  &  \multicolumn{2}{l}{ 3264} &  \multicolumn{2}{l}{2164} & \multicolumn{2}{l}{60}  \tabularnewline
  \dbline
  \end{tabular}
  }
 \end{table}

%% file: sections/Results_Letter.tex
\begin{figure}[!h]
 \centering
 \includegraphics[width=0.55\textwidth]{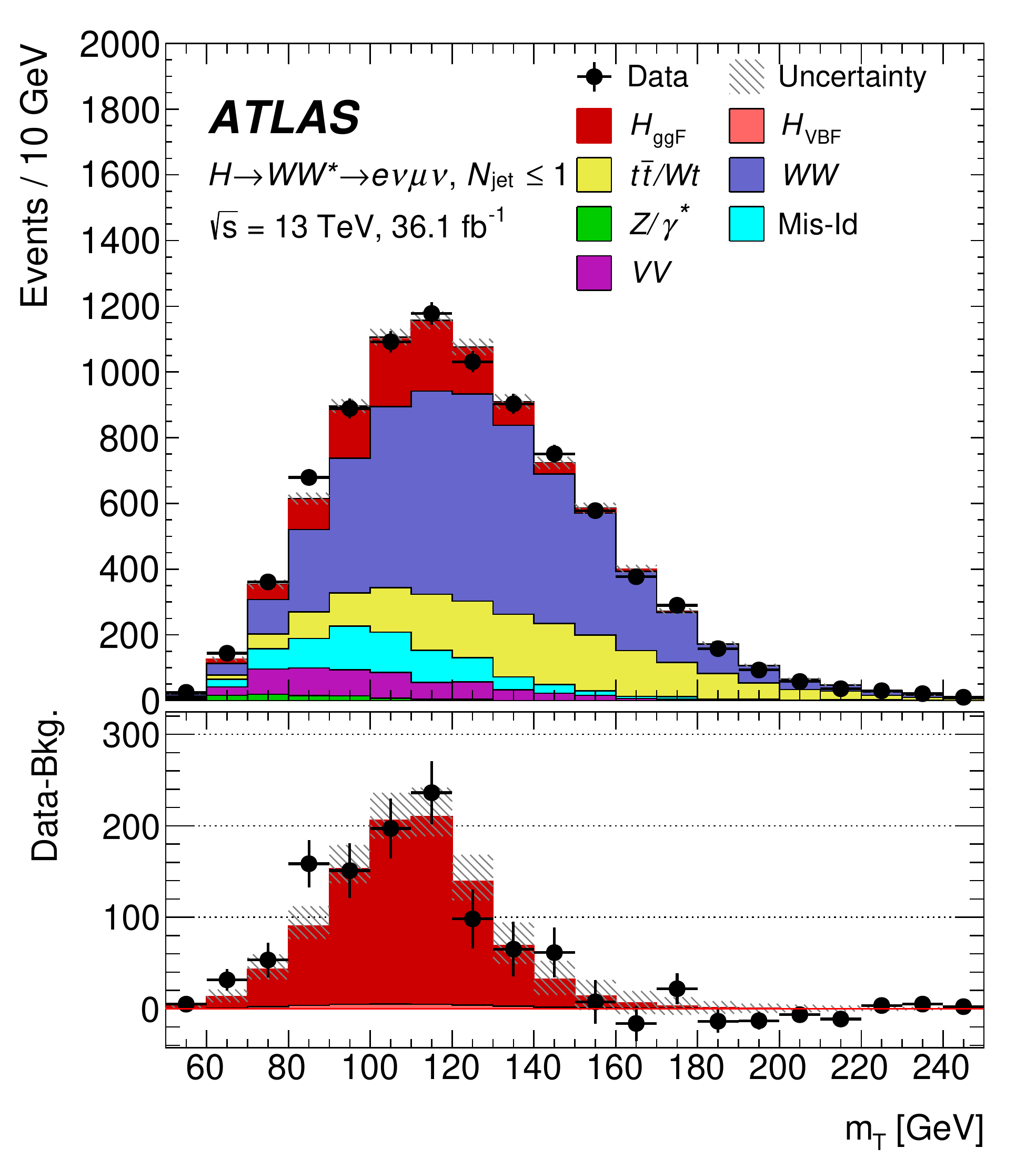}
{\caption{Post-fit combined transverse mass distribution for \ZeroOneJetSimple. 
The bottom panel shows the difference between the data and the estimated background compared to the distribution for a SM Higgs boson with $\mH = 125~\GeV$.
The signal and the background modelled contributions are fitted to the data with a floating signal strength. 
The hatched band shows the total uncertainty of the signal and background modelled contributions. 
The $H_{VBF}$ contribution is too small to be visible.
\label{fig:MT_0j1j}}}
\end{figure}
\begin{figure}[!h]
\centering
 \includegraphics[width=0.5\textwidth]{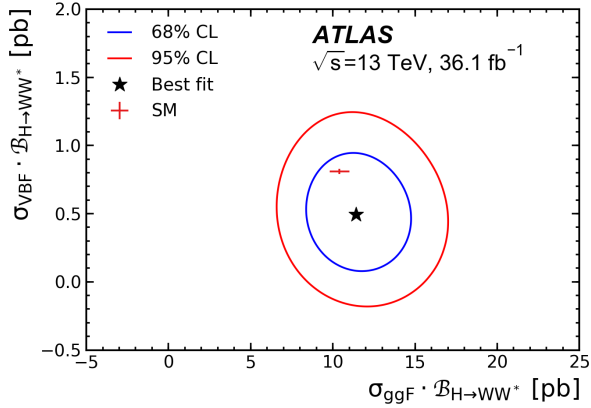}
{\caption{68\% and 95\% confidence level two-dimensional likelihood contours of $\sigma_{\mathrm{ggF}} \cdot \mathcal{B}_{\hww}$ vs. \\$\sigma_{\mathrm{VBF}}~\cdot~\mathcal{B}_{\hww}$, compared to the SM prediction shown by the red marker. The error bars on the SM prediction represent the ggF and VBF theory uncertainty~\cite{deFlorian:2016spz}, respectively.
\label{fig:LL2D}}}
\end{figure}

Figure~\ref{fig:MT_0j1j} shows the combined \mt distribution for \ZeroOneJetSimple. The bottom panel of Figure~\ref{fig:MT_0j1j} shows the difference between the data and the total estimated background compared to the $\mt$ distribution of a SM Higgs boson with $\mH = 125~\GeV$. 
The total signal observed (see Table~\ref{tab:ggF_VBF_yields}) of about 1000 events is in agreement, in both shape and rate, with the expected SM signal. 
The cross-sections times branching fractions, $\sigma_{\mathrm{ ggF}} \cdot \mathcal{B}_{\hww}$ and $\sigma_{\mathrm{VBF}} \cdot \mathcal{B}_{\hww}$, are simultaneously determined to be: 
\begin{eqnarray*}
\sigma_{\mathrm{ggF}} \cdot \mathcal{B}_{H \to WW^{\ast}} &=& 11.4^{+1.2}_{-1.1} (\mathrm{stat.})^{+1.2}_{-1.1}(\mathrm{theo~ syst.})^{+1.4}_{-1.3}(\mathrm{exp~ syst.})~\mathrm{pb}      = 11.4 ^{+2.2}_{-2.1}~\mathrm{pb} \\
\sigma_{\mathrm{VBF}} \cdot \mathcal{B}_{H \to WW^{\ast}} &=& 0.50 ^{+0.24}_{-0.22} (\mathrm{stat.}) \pm 0.10 (\mathrm{theo~ syst.})^{+0.12}_{-0.13}(\mathrm{exp~ syst.})~\mathrm{pb} = 0.50 ^{+0.29}_{-0.28}~\mathrm{pb}.
\end{eqnarray*}
The predicted cross-section times branching fraction values are $ 10.4 \pm 0.6$ pb and $0.81\pm 0.02$ pb for ggF and VBF~\cite{deFlorian:2016spz}, respectively. 
The 68\% and 95\% confidence level two-dimensional contours of $\sigma_{\mathrm{ggF}} \cdot \mathcal{B}_{H \to WW^{\ast}}$ and $\sigma_{\mathrm{VBF}} \cdot \mathcal{B}_{H \to WW^{\ast}}$ are shown in Figure~\ref{fig:LL2D} and are consistent with the SM predictions. 

The signal strength parameter $\mu$ is defined as the ratio of the measured signal yield to that predicted by the SM.
The measured signal strengths for the ggF and VBF production modes in the $\hww$ decay channel are simultaneously determined to be
\begin{eqnarray*}
\mu_{\mathrm{ggF}} &=&  1.10^{+0.10}_{-0.09} (\mathrm{stat.}) ^{+0.13}_{-0.11}(\mathrm{theo~ syst.})^{+0.14}_{-0.13}(\mathrm{exp~ syst.}) = 1.10^{+0.21}_{-0.20} \\
\mu_{\mathrm{VBF}} &=&  0.62^{+0.29}_{-0.27} (\mathrm{stat.})^{+0.12}_{-0.13} (\mathrm{theo~ syst.})\pm 0.15(\mathrm{exp~ syst.}) = 0.62^{+0.36}_{-0.35}.
\end{eqnarray*}
\input{sections/SystematicsTable_Letter.tex}\\ 
The observed (expected) ggF and VBF signals have significances of 6.0 (5.3) and 1.8 (2.6) standard deviations, respectively.


%% file: sections/SystematicsTable_Letter.tex
\begin{table}[!h]
\centering
 \caption{Breakdown of the main contributions to the total uncertainty in $\sigma_{\mathrm{ggF}} \cdot \mathcal{B}_{\hww}$ and $\sigma_{\mathrm{VBF}}  \cdot \mathcal{B}_{\hww}$. The individual sources of systematic uncertainties are grouped together. The sum in quadrature of the individual components differs from the total uncertainty due to correlations between the components.}
 \scalebox{0.8}{
\begin{tabular}{lrr}
\hline\hline
Source          & $\Delta\sigma_{\mathrm{ggF}} \cdot \mathcal{B}_{\hww}$ $[\%]$  & $\Delta\sigma_{\mathrm{VBF}} \cdot \mathcal{B}_{\hww}$ $[\%]$  \\
\hline\hline
Data statistics                                 & 10 		& 46    \\
CR statistics                                   & 7 		& 9     \\
\hline
MC statistics                                   & 6		& 21    \\
Theoretical uncertainties                       & 10         	& 19    \\
\hspace*{4mm}ggF signal                         & 5		& 13    \\
\hspace*{4mm}VBF signal                         & $<$1 		& 4     \\
\hspace*{4mm}$WW$                      		& 6		& 12    \\
\hspace*{4mm}Top-quark                          & 5		& 5     \\
Experimental uncertainties              	& 8        	& 9     \\
\hspace*{4mm}$b$-tagging      			& 4		& 6     \\
\hspace*{4mm}Modelling of pile-up         	& 5		& 2     \\
\hspace*{4mm}Jet                		& 2 		& 2     \\
\hspace*{4mm}Lepton              		& 3		& $<$1  \\
\hspace*{4mm}Misidentified leptons              & 6		& 9     \\
Luminosity                                      & 3 		& 3     \\
\hline
TOTAL                                           & 18            & 57    \\
\hline\hline
\end{tabular}
}
\label{tab:MUerrors}
\end{table}
Table~\ref{tab:MUerrors} shows the relative impact of the main uncertainties on the measured values for $\sigma_{\mathrm{ggF}}  \cdot \mathcal{B}_{\hww}$ and $\sigma_{\mathrm{VBF}}  \cdot \mathcal{B}_{\hww}$. 
The theory uncertainties in the non-resonant $WW$ background produce one of the largest uncertainties, of the order of $6\%$, in the measured ggF cross-section.
The uncertainty in the ratio of $gg \to WW$ to $qq \to WW$ comes from the limited NLO accuracy of the $gg \to WW$ production cross-section~\cite{Caola:2015rqy}. 
The resulting uncertainty in the cross-section when using acceptance criteria similar to those in this analysis was evaluated in Ref.~\cite{Melia2012} for $\ZeroJet$ and for $\OneJet$.
In the $\TwoJet$ VBF SR, the 12\% uncertainty in the $WW$ background originates from the matching and UEPS modelling of $qq \to WW$. 
The amount of ggF contamination in the VBF region is subject to QCD scale uncertainties and this produces an uncertainty of about 13\% in the measured VBF cross-section. The statistical uncertainty of the MC simulation has a relatively large 
impact, especially for the VBF cross-section measurement, where it contributes $21\%$.

%% file: sections/Conclusions_Letter.tex
Measurements of the inclusive cross-section of Higgs boson production
via the gluon--gluon fusion (ggF) and vector-boson fusion (VBF) modes in 
the \hww decay channel are presented.
They are based on 36.1~fb$^{-1}$ of $\sqrt{s} = 13$~\TeV\ proton--proton
collisions recorded by the ATLAS detector at the LHC in 2015--2016.
The ggF and VBF cross-sections times the $H \to WW^{\ast}$ branching
ratio are measured to be $11.4^{+1.2}_{-1.1} (\mathrm{stat.})
^{+1.8}_{-1.7}(\mathrm{syst.})$~pb and $0.50 ^{+0.24}_{-0.22}
(\mathrm{stat.}) \pm 0.17(\mathrm{syst.})$~pb, respectively, in
agreement with SM predictions.

%% file: sections/Acknowledgements.tex

We thank CERN for the very successful operation of the LHC, as well as the
support staff from our institutions without whom ATLAS could not be
operated efficiently.

We acknowledge the support of ANPCyT, Argentina; YerPhI, Armenia; ARC, Australia; BMWFW and FWF, Austria; ANAS, Azerbaijan; SSTC, Belarus; CNPq and FAPESP, Brazil; NSERC, NRC and CFI, Canada; CERN; CONICYT, Chile; CAS, MOST and NSFC, China; COLCIENCIAS, Colombia; MSMT CR, MPO CR and VSC CR, Czech Republic; DNRF and DNSRC, Denmark; IN2P3-CNRS, CEA-DRF/IRFU, France; SRNSFG, Georgia; BMBF, HGF, and MPG, Germany; GSRT, Greece; RGC, Hong Kong SAR, China; ISF and Benoziyo Center, Israel; INFN, Italy; MEXT and JSPS, Japan; CNRST, Morocco; NWO, Netherlands; RCN, Norway; MNiSW and NCN, Poland; FCT, Portugal; MNE/IFA, Romania; MES of Russia and NRC KI, Russian Federation; JINR; MESTD, Serbia; MSSR, Slovakia; ARRS and MIZ\v{S}, Slovenia; DST/NRF, South Africa; MINECO, Spain; SRC and Wallenberg Foundation, Sweden; SERI, SNSF and Cantons of Bern and Geneva, Switzerland; MOST, Taiwan; TAEK, Turkey; STFC, United Kingdom; DOE and NSF, United States of America. In addition, individual groups and members have received support from BCKDF, CANARIE, CRC and Compute Canada, Canada; COST, ERC, ERDF, Horizon 2020, and Marie Sk{\l}odowska-Curie Actions, European Union; Investissements d' Avenir Labex and Idex, ANR, France; DFG and AvH Foundation, Germany; Herakleitos, Thales and Aristeia programmes co-financed by EU-ESF and the Greek NSRF, Greece; BSF-NSF and GIF, Israel; CERCA Programme Generalitat de Catalunya, Spain; The Royal Society and Leverhulme Trust, United Kingdom. 

The crucial computing support from all WLCG partners is acknowledged gratefully, in particular from CERN, the ATLAS Tier-1 facilities at TRIUMF (Canada), NDGF (Denmark, Norway, Sweden), CC-IN2P3 (France), KIT/GridKA (Germany), INFN-CNAF (Italy), NL-T1 (Netherlands), PIC (Spain), ASGC (Taiwan), RAL (UK) and BNL (USA), the Tier-2 facilities worldwide and large non-WLCG resource providers. Major contributors of computing resources are listed in Ref.~\cite{ATL-GEN-PUB-2016-002}.

%% file: atlas_authlist.tex
 
\begin{flushleft}
{\Large The ATLAS Collaboration}

\bigskip

M.~Aaboud$^\textrm{\scriptsize 34d}$,    
G.~Aad$^\textrm{\scriptsize 99}$,    
B.~Abbott$^\textrm{\scriptsize 124}$,    
O.~Abdinov$^\textrm{\scriptsize 13,*}$,    
B.~Abeloos$^\textrm{\scriptsize 128}$,    
D.K.~Abhayasinghe$^\textrm{\scriptsize 91}$,    
S.H.~Abidi$^\textrm{\scriptsize 164}$,    
O.S.~AbouZeid$^\textrm{\scriptsize 39}$,    
N.L.~Abraham$^\textrm{\scriptsize 153}$,    
H.~Abramowicz$^\textrm{\scriptsize 158}$
% ...
\bigskip
\\

% ...
$^{13}$Institute of Physics, Azerbaijan Academy of Sciences, Baku; Azerbaijan.\\
% ...
$^{34}$$^{(a)}$Facult\'e des Sciences Ain Chock, R\'eseau Universitaire de Physique des Hautes Energies - Universit\'e Hassan II, Casablanca;$^{(b)}$Centre National de l'Energie des Sciences Techniques Nucleaires (CNESTEN), Rabat;$^{(c)}$Facult\'e des Sciences Semlalia, Universit\'e Cadi Ayyad, LPHEA-Marrakech;$^{(d)}$Facult\'e des Sciences, Universit\'e Mohamed Premier and LPTPM, Oujda;$^{(e)}$Facult\'e des sciences, Universit\'e Mohammed V, Rabat; Morocco.\\
% ...
$^{39}$Niels Bohr Institute, University of Copenhagen, Copenhagen; Denmark.\\
% ...
$^{91}$Department of Physics, Royal Holloway University of London, Egham; United Kingdom.\\
% ...
$^{99}$CPPM, Aix-Marseille Universit\'e, CNRS/IN2P3, Marseille; France.\\
% ...
$^{124}$Homer L. Dodge Department of Physics and Astronomy, University of Oklahoma, Norman OK; United States of America.\\
% ...
$^{128}$LAL, Universit\'e Paris-Sud, CNRS/IN2P3, Universit\'e Paris-Saclay, Orsay; France.\\
% ...
$^{153}$Department of Physics and Astronomy, University of Sussex, Brighton; United Kingdom.\\
% ...
$^{158}$Raymond and Beverly Sackler School of Physics and Astronomy, Tel Aviv University, Tel Aviv; Israel.\\
% ...
$^{164}$Department of Physics, University of Toronto, Toronto ON; Canada.\\
% ...
$^{a}$ Also at Borough of Manhattan Community College, City University of New York, NY; United States of America.\\
$^{b}$ Also at Centre for High Performance Computing, CSIR Campus, Rosebank, Cape Town; South Africa.\\
$^{c}$ Also at CERN, Geneva; Switzerland.\\
$^{d}$ Also at CPPM, Aix-Marseille Universit\'e, CNRS/IN2P3, Marseille; France.\\
$^{e}$ Also at D\'epartement de Physique Nucl\'eaire et Corpusculaire, Universit\'e de Gen\`eve, Gen\`eve; Switzerland.\\
% ...
$^{*}$ Deceased

\end{flushleft}
